\documentclass[referee]{raa_rb}           

\usepackage{graphicx,times}
\usepackage{natbib}
\usepackage{subcaption}
\usepackage{amssymb,amsmath}
\usepackage{enumitem}
\bibpunct{(}{)}{;}{a}{}{,}
\usepackage[pagebackref=true]{hyperref}
\usepackage{lmodern}

\usepackage{tikz}
\usetikzlibrary{positioning}
\usetikzlibrary{shapes,arrows,arrows,positioning,fit}
\usepackage[symbol]{footmisc}

\begin{document}
   \title{Mock Observations for the CSST Mission: Main Surveys--An Overview of Framework and Simulation Suite}
   \volnopage{ {\bf 20XX} Vol.\ {\bf X} No. {\bf XX}, 000--000}
   \setcounter{page}{1}

   \author{Cheng-Liang Wei\inst{1,*}, Guo-Liang Li\inst{1,*}\footnotetext{$*$ These authors contributed equally to this work.}, Yue-Dong Fang\inst{2,}$^\dagger$\footnotetext{$\dagger$ Corresponding Authors.}, Xin Zhang\inst{3}, Yu Luo\inst{4}, Hao Tian\inst{3}, De-Zi Liu\inst{5}, Xian-Ming Meng\inst{3}, Zhang Ban\inst{6}, Xiao-Bo Li\inst{6}, Zun Luo\inst{3}, Jing-Tian Xian\inst{6}, Wei Wang\inst{6}, Xi-Yan Peng\inst{7},  Nan Li\inst{3}, Ran Li\inst{3,8},  Li Shao\inst{3}, Tian-Meng Zhang\inst{3}, Jing Tang\inst{3}, Yang Chen\inst{9}, Zhao-Xiang Qi\inst{7}, Zi-Huang Cao\inst{3},  Huan-Yuan Shan\inst{7}, Lin Nie\inst{7}, Lei Wang\inst{1}, Zizhao He\inst{1}, Rui-Biao Luo\inst{1}, Quan-Yu Liu\inst{1}, Zhaojun Yan\inst{7}}

    \institute{Purple Mountain Observatory, Chinese Academy of Sciences, 10 Yuanhua Road, Nanjing 210023, People’s Republic of China;
	\and Universit\"ats-Sternwarte M\"unchen, Fakult\"at f\"ur Physik, Ludwig-Maximilians-Universit\"at M\"unchen, Scheinerstrasse 1, 81679 M\"unchen, Germany; {\it \textbf{Yuedong.Fang@physik.uni-muenchen.de}}
 	\and National Astronomical Observatories, Chinese Academy of Sciences, 20A Datun Road, Beijing 100101, People’s Republic of China;
        \and School of Physics and Electronics, Hunan Normal University, 36 Lushan Road, Changsha 410081,People’s Republic of China;
	\and South-Western Institute for Astronomy Research, Yunnan University, Kunming, Yunnan, 650500, People’s Republic of China
        \and{Space Optics Department, Changchun Institute of Optics, Fine Mechanics and Physics, Chinese Academy of Sciences, Changchun, 130033, People’s Republic of China;}
	\and Shanghai Astronomical Observatory, Chinese Academy of Sciences, Shanghai 200030, People’s Republic of China
	\and School of Physics and Astronomy, Beijing Normal University, Beijing 100875, People’s Republic of China
	\and School of Physics and Optoelectronic Engineering, Anhui University, Hefei, 230601, People’s Republic of China
	}

\abstract{The Chinese Space Station Survey Telescope (CSST) is a flagship space-based observatory. Its main survey camera is designed to conduct high spatial resolution near-ultraviolet to near-infrared imaging and low-resolution spectroscopic surveys. To maximize the scientific output of CSST, we have developed a comprehensive, high-fidelity simulation pipeline for reproducing both imaging and spectroscopic observations. This paper presents an overview of the simulation framework, detailing its implementation and components. Built upon the GalSim package and incorporating the latest CSST instrumental specifications, our pipeline generates pixel-level mock observations that closely replicate the expected instrumental and observational conditions. The simulation suite integrates realistic astrophysical object catalogs, instrumental effects, point spread function (PSF) modeling, and observational noises to produce accurate synthetic data. We describe the key processing stages of the simulation, from constructing the input object catalogs to modeling the telescope optics and detector responses. Furthermore, we introduce the most recent release of simulated datasets, which provide a crucial testbed for data processing pipeline developments, calibration strategies, and scientific analyses, ensuring that CSST will meet its stringent requirements. Our pipeline serves as a vital tool for optimizing CSST main survey strategies and ensuring robust cosmological measurements.
\keywords{software: simulations --- software: public release --- methods: numerical ---  surveys --- virtual observatory tools}
}
   \authorrunning{C.-L. Wei et al. }            
   \titlerunning{An overview of CSST synthetic observation}  
   \maketitle

\section{Introduction}           
\label{sect:intro}

The field of cosmology has undergone a remarkable transformation in recent decades, evolving from a qualitative and theoretical discipline into a highly precise and data-driven science. This shift, often referred to as the era of precision cosmology, has been fueled by advances in observational techniques, data analysis methodologies, and theoretical modeling. High-precision measurements of cosmic structures, temperature fluctuations in the cosmic microwave background (CMB), large-scale structure distributions, and Type Ia supernovae have allowed cosmologists to place stringent constraints on fundamental cosmological parameters \citep{Riess_1998, Planck_2018}. A key milestone in precision cosmology was the establishment of the $\Lambda$CDM model as the standard cosmological paradigm. This framework successfully explains a broad range of cosmological phenomena, including the accelerated expansion of the universe, the formation and evolution of large-scale structures, and the abundances of light elements predicted by Big Bang nucleosynthesis.

In this new era of precision cosmology, stage IV surveys represent the next generation of large-scale cosmic observations, designed to push the boundaries of our understanding. These surveys, characterized by their deep, wide-area coverage and high-precision measurements, aim to tackle key cosmological challenges such as the nature of dark energy and the expansion history of the  Universe. Examples of Stage IV surveys include space-based missions like the Chinese Space Station Survey Telescope (CSST), Euclid (\citealt{Euclid_Definition, Euclid_prepare}), and the Nancy Grace Roman Space Telescope (\citealt{Roman_report}), each of  them brings unique capabilities to this global effort. Among them, the CSST is designed to conduct a high spatial resolution near-ultraviolet to  near-infrared imaging and low-resolution spectroscopic surveys of the  Universe. It is expected to deliver deep, high-precision observations for a broad range of astrophysical studies, including cosmology, galaxy evolution, stellar astrophysics, and exoplanet research \citep{Zhan2021, Gong_2025}.

The increasing precision of cosmological measurements has, however, introduced new challenges. Systematic uncertainties, observational biases, and theoretical degeneracies must be rigorously controlled to ensure that the inferred cosmological parameters are robust. Realizing the full statistical potential of Stage IV surveys necessitates stringent control over various systematic uncertainties. In weak lensing cosmological studies, for example, the multiplicative shear measurement bias must be constrained to an accuracy of approximately $10^{-3}$ \citep{2013MNRAS.429..661M}. Similarly, the spectroscopic redshift bias and photometric redshift (photo-z) scatter across different redshift bins should remain below $10^{-3}(1 + z)$ and $0.05(1 + z)$, respectively \citep{Euclid_Definition}. Given these stringent requirements, we developed a realistic and comprehensive imaging simulation pipeline, which is essential for maximizing the scientific return of the CSST Survey Camera (SCam). This pipeline is designed to: 
\begin{itemize}
    \item Assess the overall performance of the instrument, accounting for factors such as telescope optics, detector noise, and variations in the point spread function (PSF).
    \item Validate data processing pipelines by providing extensive test datasets, incorporating diverse ground truths and operational conditions.
    \item Provide standardized mock datasets to support other key scientific objectives. 
    \item Provide user-friendly simulation software, enabling researchers with specific scientific needs to independently run small-scale, customized simulations.
\end{itemize}
This simulation pipeline is built on the GalSim~\citep{Rowe2015} package, integrating the latest CSST instrumental specifications and lab-measured characteristics. It generates high-fidelity, pixel-level mock data for both imaging and spectroscopic observations of CSST.

This paper is organised as follows: Section~\ref{sect:mission} provides an overview of the CSST mission and its onboard scientific instruments. We explain how we construct the object catalogs, which serve as the ground truth inputs for our simulations in Section~\ref{sect:TU}. Next, we give a general introduction of the workflow and highlights key processing stages of the simulation pipeline in Section~\ref{sect:workflow}. In Section~\ref{sect:suite}, we describe in detail how we model the CSST Point-spread Function (PSF), the astronomical objects, and various instrumental effects. Section~\ref{sect:products} presents an overview of the Cycle-9 released simulated data product which is generated by our pipeline. Finally, we summarize and conclude our results in Section~\ref{sect:summary}.

\section{Mission Overview}
\label{sect:mission}
\begin{figure}
    \centering
    \includegraphics[width=0.75\linewidth]{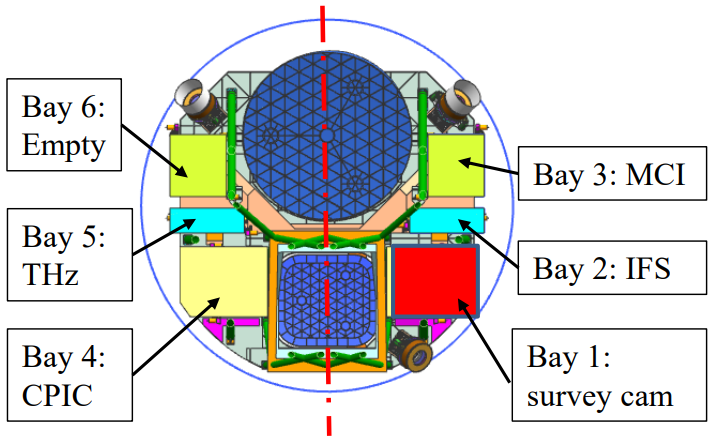}
    \caption{The arrangement of the currently planed scientific instrument terminals for CSST mission.}
    \label{fig:all_terminals}
\end{figure}
The China Manned Space Station (CMS), also known as Tiangong, is a permanently crewed space station built and operated by the China Manned Space Agency. The station consists of three main modules and is assembled in a low-Earth orbit at an altitude of 340 to 450 km. It is designed to support the long-term human presence in space and provides a platform for conducting space-based scientific research.

The Chinese Space Station Survey Telescope (CSST), also referred to as the Xuntian Space Telescope, is an independent optical space observatory designed to operate in the same orbit as the  CMS (\citealt{Zhan2011}, \citealt{Gong2019}, \citealt{Zhan2021}). It can dock with the station for servicing when needed.
The CSST features a single primary optical system with a 2-meter aperture $f/14$ Cooke-type off-axis three-mirror anastigmat  design, which collects light from observed targets. By utilizing a folding mirror near the exit pupil, the system can switch between different instruments at its terminals to capture images, spectra, and other data, respectively.
. The first-generation observation terminals, to be launched with the telescope into orbit, include five instruments: Multi-band Imaging and Slitless Spectroscopy Survey Camera (also known as the Survey Camera, SCam), Multi-Channel Imager (MCI), Integral Field Spectrograph (IFS), Cool Planet Imaging Coronagraph (CPI-C), and THz Spectrometer (TS). The arrangement of these instrument terminals is shown in  Fig.~\ref{fig:all_terminals}.

\begin{figure*}
    \centering
    \includegraphics[width=1.\linewidth]{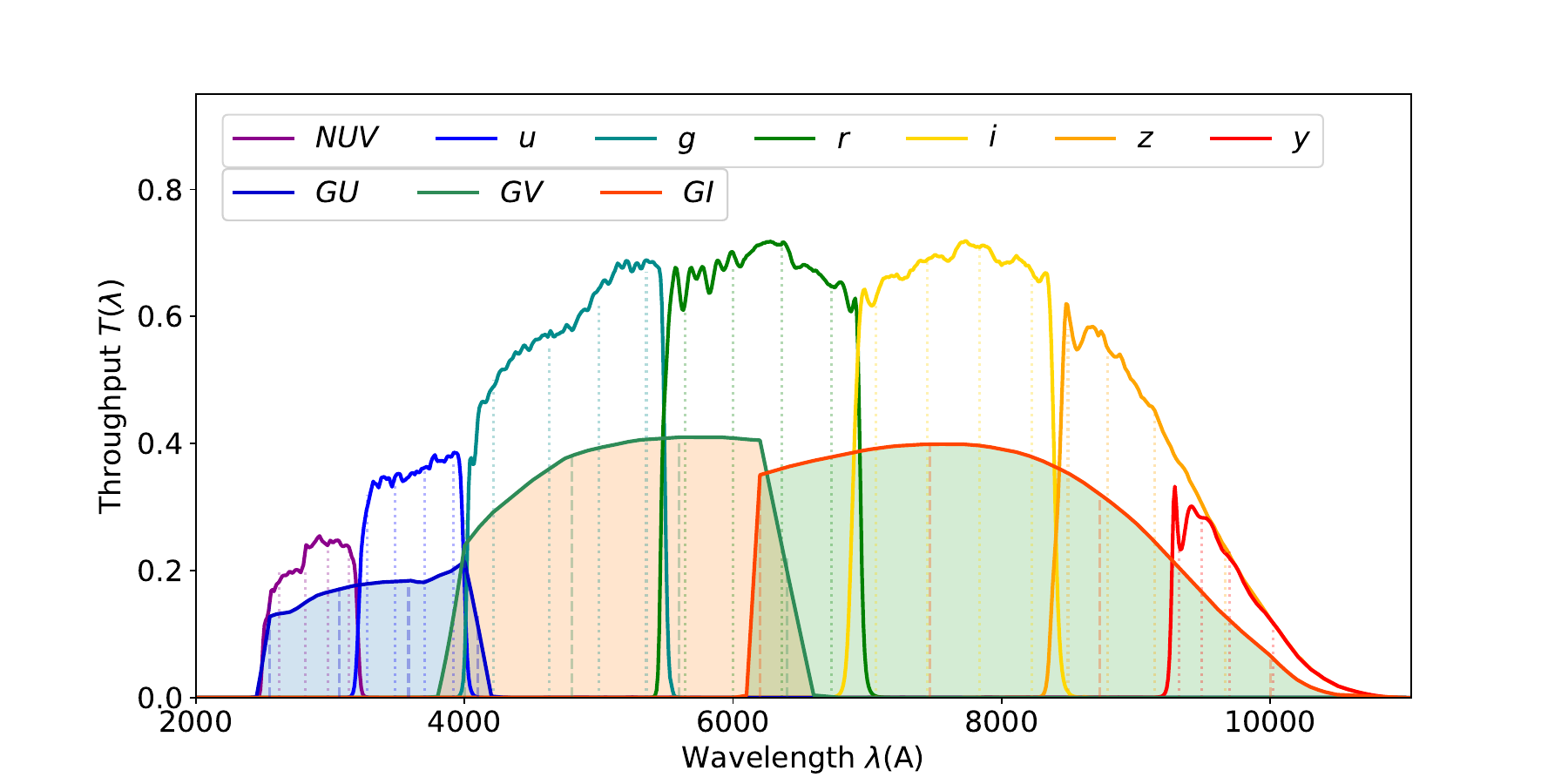}
    \caption{Response curves of the CSST passbands, accounting for the combined efficiencies of the mirrors, filters, and detector quantum efficiency (QE). The vertical dashed lines indicate the specific wavelengths at which the PSFs are sampled in our simulation.}
    \label{fig:throughputs}
\end{figure*}

\begin{figure}
    \centering
    \includegraphics[width=0.75\textwidth]{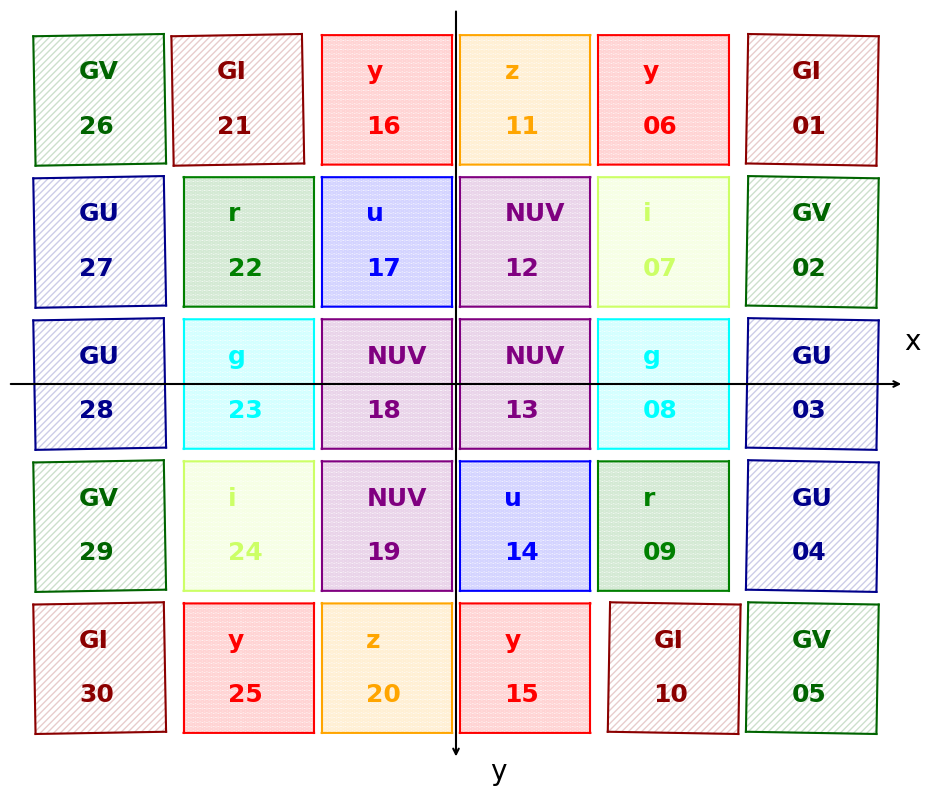}
    \caption{Focal plane layout of the CSST main survey camera (SCam). The array consists of 30 detectors arranged in a 6 × 5 grid, with each detector equipped with either a filter or a filter combined with a grism. The central detectors are dedicated to photometric observations across the \textit{NUV, u, g, r, i, z}, and \textit{y} bands, while the detectors along the edges are assigned to spectroscopic observations in the \textit{GU, GV}, and \textit{GI} bands. The shutter opens along the x-axis.}
    \label{fig:focal_plane_layout}
\end{figure}

 \textbf{The SCam} is designed to carry out the primary mission of the CSST, performing multi-band imaging and slitless spectroscopic surveys. The majority of the observing time will be allocated to SCam. Its scientific observation focal plane consists of 2.6 billion pixels, corresponding to a field of view of 1.1 square degrees, with a dynamic image quality of EE80$ \leq 0.15''$ (radii of the PSF at which 80\% of the energy is encircled). The focal plane is divided into seven imaging bands (\textit{NUV, u, g, r, i, z, y}) and three slitless spectroscopic bands (\textit{GU, GV, GI}) with an average spectral resolution of $\geq 200$, covering  a wavelength range of $255–1000$ nm. The bandpass response curves are shown in  Fig.~\ref{fig:throughputs}. The primary focal plane of the survey camera is shown in Fig.~\ref{fig:focal_plane_layout}.

The plan for the telescope's ten-year mission includes approximately seven years of cumulative survey observation. This will yield wide-field data covering 17,500 square degrees in seven imaging bands and three low-resolution spectroscopic bands, along with deep-field observations over 400 square degrees in the same bands. The CSST wide-field survey will primarily cover regions at medium to high Galactic and ecliptic latitudes. It is expected to achieve a g-band limiting magnitude of 26 (for point sources, $5 \sigma $, AB magnitude), collecting photometric data for more than 1 billion galaxies and 1 billion stars, along with millions of spectra. This dataset will form the cornerstone for cutting-edge research with the CSST and will provide valuable observational targets for the other four terminal instruments as well as other astronomical facilities.

 \textbf{The MCI} employs a relay optical system to extend the focal length, achieving a higher image sampling rate. It uses a dichroic beam splitter to split the light path of its $7.5' \times 7.5'$  field of view into three wavelength channels: near-ultraviolet, blue, and red, allowing simultaneous imaging in all three bands. This setup facilitates the precise capture of transient source states. Each channel is equipped with 10 filters, catering to a wide range of observational needs, from studies within the Solar System to those of the distant Universe.

The core scientific goals of the MCI include establishing a high-precision standard star catalog and conducting ultra-deep field observations in ultraviolet and visible bands. The former ensures high-precision photometry for the CSST survey, while the latter provides unprecedented depth in ultraviolet and visible observations. These capabilities will play a critical role in studies of galaxy and black hole co-evolution, deep-field science involving strongly lensed galaxy clusters, Type Ia supernova cosmology, asteroids, and rapidly changing celestial phenomena.

 \textbf{The IFS} uses a slicer to divide a field of view of no less than $6''\times 6''$ into multiple $0.2''\times 6''$ units, which are then separated by a dichroic beam splitter and rearranged for spectral dispersion. This process covers two wavelength channels ( red: $350-580$ nm, and blue $580-1000$ nm), achieving a spatial resolution of approximately $0.2''$ and a spectral resolution of $R \geq 1000$. The IFS enables imaging spectroscopy observations, simultaneously capturing both the two-dimensional structure and spectral information of the observed targets.

The IFS is well-suited for studies requiring spatially resolved analysis of the chemical composition or physical properties of targets. Applications include research on the physical properties of the nuclear region around supermassive black holes in galaxy centers, the co-evolution of galaxies and black holes, star formation under specific galactic conditions, the dynamics and dark matter distribution of strongly lensed galaxies, Galactic nebulae, tidal disruption events involving black holes, and objects within the Solar System.

 \textbf{The CPI-C} employs pupil modulation and high-precision wavefront error correction technologies to suppress diffraction photon noise and speckle noise generated by the primary optical system. This enables ultra-high contrast imaging of exoplanets $(\leq 10^{-8}$ at 600–900 nm, with an inner working angle of no more than $0.55''$ at 633 nm).

The coronagraph is designed to search for Jupiter-like planets and super-Earths located near the habitable zones or snow lines (0.8–5 AU) of nearby  Solar-type stars. It aims to directly image these exoplanets with ultra-high contrast and conduct multi-band photometric studies. Additionally, the instrument will perform high-contrast imaging and quantitative analysis of circumstellar disks and exozodiacal dust around stars, providing critical observational evidence to advance theories on planetary formation and evolution.

 \textbf{The TS} uses niobium nitride-based superconductor-insulator-superconductor  mixer technology to receive and mix signals in the 0.41–0.51 THz frequency range collected by the primary optics. Operating at an ultra-low temperature of approximately 8K, it produces spectral line data with a frequency resolution of  $\leq 100 \, \text{kHz}$, enabling high-sensitivity detection of terahertz signals from celestial objects.

The TS will be used for spectral line surveys and CI mapping observations to investigate the chemical composition of celestial bodies and the interstellar medium, search for new molecular species, and reveal atomic-to-molecular phase transitions in galaxy evolution. It will also  get the data for studying the formation and evolution mechanisms of molecular clouds.
 By combining  the data  observed with other wavelengths, the TS will deepen our understanding  on the structural formation, dynamical evolution, chemical processes, and star formation activities in nearby galaxies.

\section{The True Universe}
\label{sect:TU}
\subsection{Stellar Catalogues}
Field stars are essential targets in the survey, as their luminosity distribution function (LDF) and number density distribution (NDD) directly influence the accuracy of results produced by the subsequent data processing pipeline.
To generate a simulated stellar catalog with realistic LDF and NDD characteristics, we adopt the Galaxia tool
\citep{Galaxia2011ApJ...730....3S}. Galaxia is an efficient and flexible framework that samples stars from the Besancon analytical model of the Milky Way \citep{Robin2003A&A...409..523R}, providing reliable predictions across visible and near-infrared photometric bands, and reproducing rotation curves consistent with Hipparcos and other observational results.
Galaxia$^{[1]}$ \footnotetext{[1] \url{https://galaxia.sourceforge.net/Galaxia3pub.html}} allows users to customize the photometric system, color–magnitude limits, sky coverage, and stellar population parameters.

For the CSST simulation, the output catalog includes stars from the thin disk, thick disk, halo, and bulge (in selected directions). However, it currently excludes globular clusters, dwarf galaxies, and stellar streams, which will be incorporated in future updates. Additionally, only single stars are included in the present stage of  simulation.
\begin{figure}
    \centering
    \includegraphics[width=0.75\textwidth]{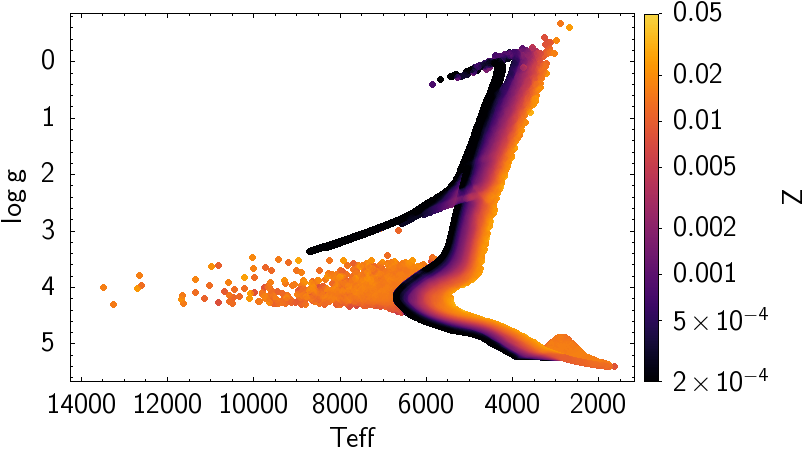}
    \caption{The distribution in the Kiel Diagram 
    of the stars in the output catalog.}
    \label{fig:StarCatHRD}
\end{figure}
 Fig.~\ref{fig:StarCatHRD} shows the distribution
of one of the input catalogs in the Kiel diagram, $\mathrm{log}\,g$ versus $T_{\mathrm{eff}}$
color-coded by metallicity. This catalog spans a wide range of stellar types, with effective temperatures ranging from $1950\, K$ to $\sim14000\,K$. 
 
In the simulation, the stellar spectra are essential for both multi-band photometric and slitless spectroscopic surveys.
 The spectra of individual stars are generated using the BT Settl library$^{[2]}$ \footnotetext{[2] \url{http://astro.vaporia.com/start/btsettl.html}} \citep{Hauschildt1997ApJ...483..390H,Allard2011ASPC..448...91A} based on the  physical parameters of the star: effective temperature $T_{\mathrm{eff}}$, surface gravity $\mathrm{log\,}g$, metallicity $\mathrm{[Fe/H]}$, and age $\tau$.
To minimize storage demands, only the stellar parameters are stored. These parameters are then used to dynamically generate spectra during the simulation.

Additionally, the instantaneous position of each star at the time of exposure must be computed, which requires its sky coordinates, distance, proper motion, and radial velocity. In the current version of the simulator, all of this astrometric information is assumed to be provided in the J2000 epoch.

\subsection{Galaxy Catalogues}
The input galaxy catalogue should freature a wide range of realistic properties to support the development of the imaging simulator and data processing. We generated the mock galaxy catalogue by a semi-analytical galaxy formation model based on the high-precision JiuTian N-body simulation \citep{Jiutian_2025arXiv250321368H}. Gravitational lensing effects for each galaxy were calculated through a curved multi-plane ray-tracing simulation of weak lensing. This section provides an overview of the mock galaxy catalogue data products used in CSST main survey simulation  and the details are available in Wei et al.(2025, accepted).

\subsubsection{Jiu-Tian cosmological simulation}
`JiuTian' simulations are a suit of $\Lambda$CDM cosmological N-body simulations with three different box sizes ($ 300h^{-1}{\rm Mpc}$, $1 h^{-1}{\rm Gpc}$ ,$1.5h^{-1}{\rm Mpc}$ ) and mass resolutions ($1.005\times 10^7 h^{-1} \rm{M}_\odot$, $3.723 \times 10^8 h^{-1} \rm{M}_\odot$,$2.978 \times 10^8 h^{-1} \rm{M}_\odot$), specifically designed for CSST (See \citealt{Jiutian_2025arXiv250321368H} for more details). JT1G, one of the `JiuTian' simulation offers both large volume ($1 h^{-1}{\rm Gpc}$) and high resolution ($3.723 \times 10^8 h^{-1} \rm{M}_\odot$), is utilzed to generate the halo catalogue using SUBFIND  algorithm \citep{Springel+2001MNRAS_subfind} and build merger trees for the semi-analytical model. The cosmological parameters follow Planck2020: $\Omega_{\rm m}=0.3111$, $\Omega_\Lambda=0.6889$, $\sigma_8=0.8102$ and $H_0=67.66\ {\rm km s^{-1} Mpc^{-1}}$ \citep{Aghanim+2020A&A_Planck}.

\subsubsection{Semi-analytical modeling}
\begin{figure*}
    \centering 
    \includegraphics[width=0.9\textwidth]
    {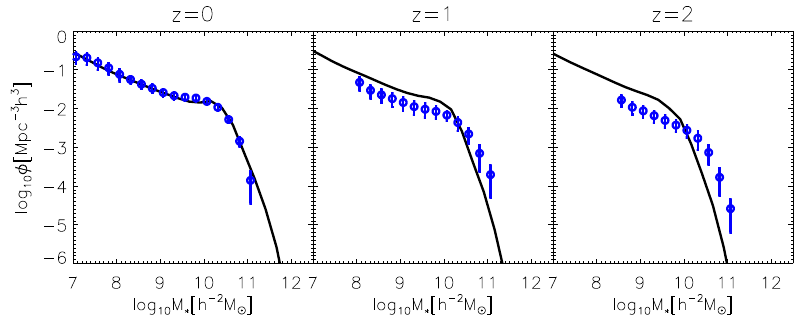}
    \caption{The evolution of stellar mass functions from $z=0$ to 2. The blue symbols are the  observation data from \cite{Henriques+2015MNRAS}. }
    \label{fig:smf-z}
\end{figure*}

Semi-analytical model (SAM)  is a powerful tool for rapidly generating mock galaxy catalogs by applying physically motivated prescriptions for galaxy formation to halo merger trees derived from N-body cosmological simulations. In this project, we adopt the SAM developed by \cite{Luo+2016MNRAS}, which builds upon the models of \cite{Guo+2013MNRAS} and \cite{Fu+2013MNRAS}, collectively known as the Munich semi-analytical model, or L-Galaxies. \cite{Luo+2016MNRAS} introduces key improvements for low-mass systems, particularly satellite galaxies, by incorporating additional physics such as cold gas stripping and an analytical treatment of orphan galaxies.

The SAM provides a comprehensive set of galaxy properties, including spatial position, redshift, velocity, dark matter halo mass and shape, stellar mass (separated into disk and bulge components), cold and hot gas masses, supermassive black hole (SMBH) mass, metallicity, and star formation history,  and others. To ensure consistency with observations, we calibrate a subset of free parameters in the model to reproduce the observed stellar mass function at redshift $z=0$.  Fig.~\ref{fig:smf-z} presents the resulting stellar mass function evolution from $z=2$ to 0.

To generate realistic mock galaxy images, we must assign each galaxy as a morphological profile and a spectral energy distribution (SED) based on its individual physical properties.
Galaxy sizes are determined using the mass–size relation derived from SDSS galaxies by \cite{Zhang+2019RAA} , with an added redshift-dependent scaling to account for cosmic size evolution.
For SED generation, we utilize STARDUSTER \citep{Qiu+2022ApJ} with star formation history and galaxy geometry parameters to generate the high-resolution SEDs. STARDUSTER is a supervised deep learning model trained on the SKIRT radiative transfer simulation \citep{Camps+2015A&C_SKIRT} which can represent features of dust attenuation and emission of a galaxy. To reduce data volume, the generated SEDs are compressed using principal component analysis (PCA). Each galaxy's rest-frame SED, sampled uniformly \del {at} with a resolution of $0.29 {\rm nm}$ over the $60{\rm nm} \sim 1100 {\rm nm}$ range, is stored using just 20 PCA coefficients, which are sufficient for accurate reconstruction via the first 20 principal components.

Quasar catalogs are generated alongside the galaxy catalog to enable joint simulations. We use SIMQSO$^{[3]}$ \footnotetext{[3] \url{https://simqso.readthedocs.io/en/latest}} \citep{McGreer+2021ascl.soft06008M} to generate quasar SEDs. SIMQSO is a suite of tools that generates mock quasar spectra based on a broken power-law continuum model and Gaussian emission line templates. We  calculated the total number of quasars in a target sky area using observed quasar luminosity functions \citep{Ross+2013ApJ} in various redshift bins. Then we generate the SEDs of individual quasars based on their luminosities. Finally, we assign these quasars to galaxies in the same redshift bins and match their luminosities to the AGN accretion rates of those galaxies from the galaxy catalog.

\begin{figure*}
    \centering
    \includegraphics[width=0.4\textwidth, angle=0]{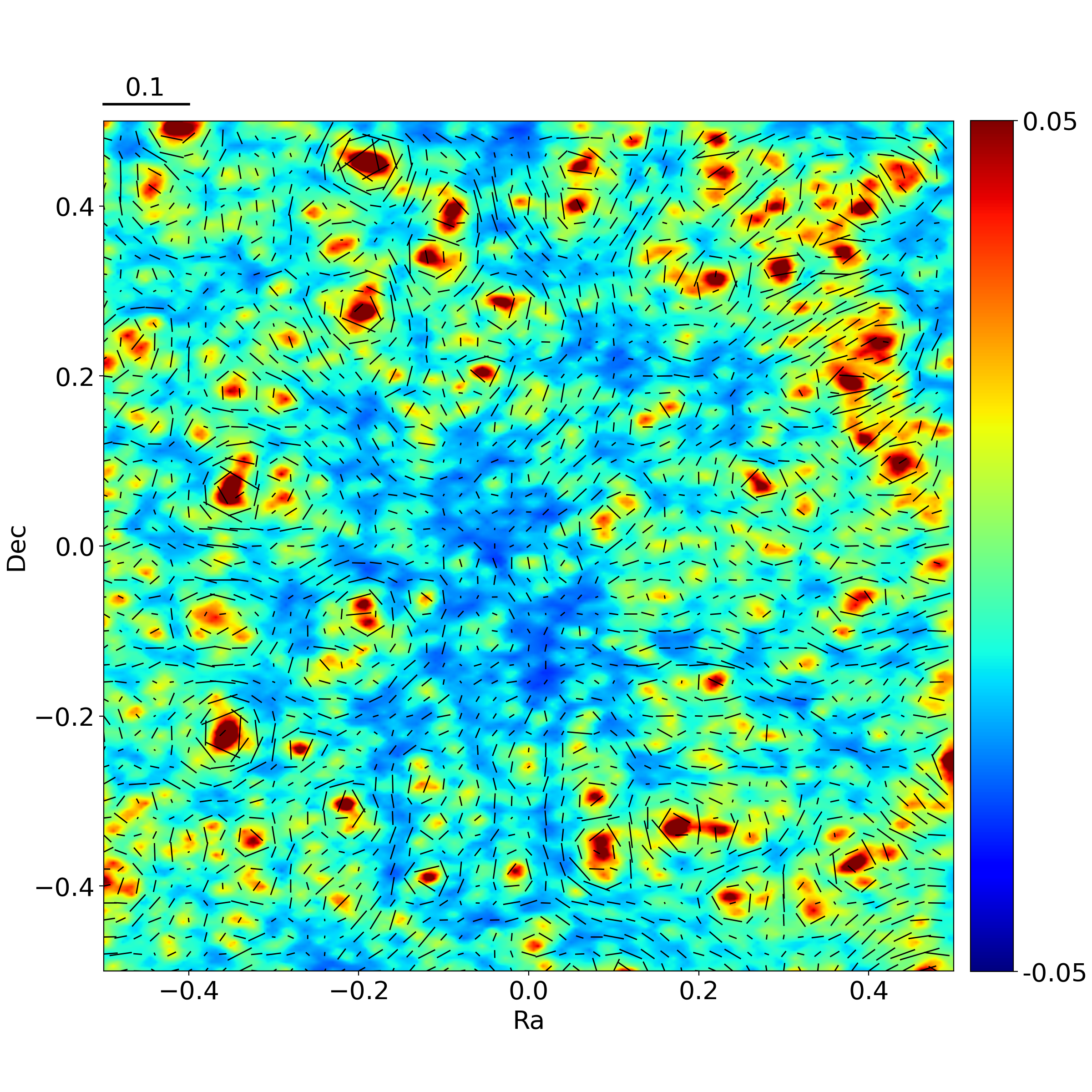}
    \raisebox{2mm}{\includegraphics[width=0.525\textwidth, angle=0]{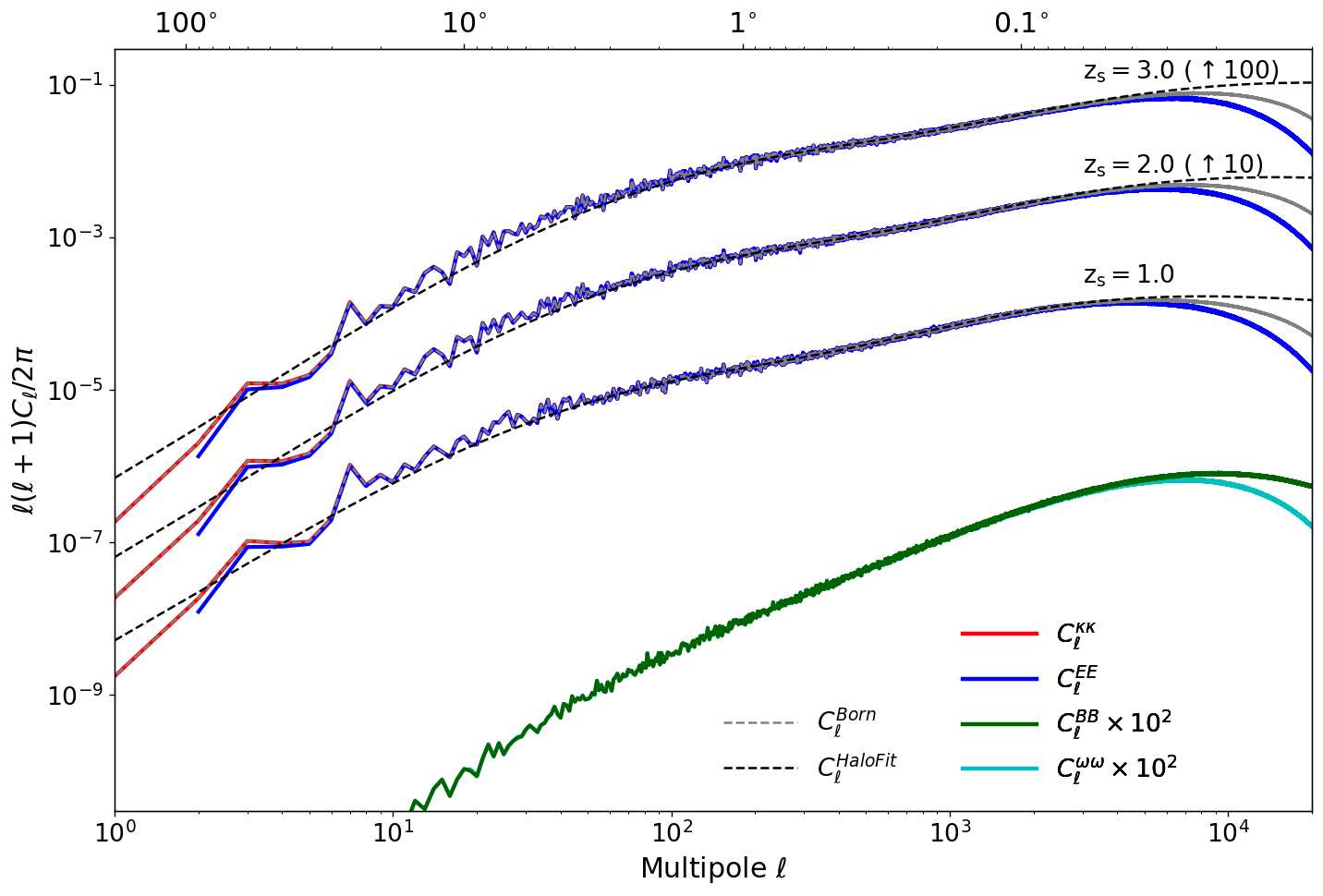}}
    \caption{Left panel shows the weak lensing maps of convergence and shear field for sources at $z_{\rm s} =1.0$ in a field of $1 \times 1 \deg^2$ area. Some tangential shears can be found around convergence peaks.
    Right panel shows the power spectrum of lensing convergence and shear for source located at different redshifts 1.0, 2.0 and 3.0, respectively.}
    \label{fig:lensing_map}
\end{figure*}

\subsubsection{Weak lensing distortions}
In the  Universe, galaxy shapes can suffer some coherent distortions caused by the intervening mass field of large-scale structures. Using the simulation of JT1G, we constructed a cosmological light cone from $z = 0$ to $z_{\rm max} = 3.5$ by stacking a sequence of cubic boxes to fully trace the observer’s past light cone. To model weak gravitational lensing effects, we performed a full-sky ray-tracing simulation, which provides lensing properties for the mock galaxy catalogue \citep{Wei+2018ApJ}. The resolution adopted in the simulation of weak lensing is 0.43 arcmin, enabling consistent calculation of weak lensing  observations down to arcminute scales. For each galaxy in the light cone, we derived the reduced shear ${\bm g} = {\bm \gamma}/(1-\kappa)$ and magnification $\mu = 1/(1-\kappa -|{\bm \gamma|})(1-\kappa+|{\bm \gamma|})$, where $\kappa$ and ${\bm \gamma} = \gamma_1 + \mathrm{i} \gamma_2$ denotes convergence and lensing shear, respectively. For more details we refer the reader to Wei et al.(2025, under review). 

 Fig.~\ref{fig:lensing_map} (left panel) presents an example of the convergence map within a patch field of $1 \times 1 \deg^2$, where the clumpy regions are typically dominated by very massive haloes \citep{wei+2018MNRAS}. In these regions, galaxies are tangentially stretched, as illustrated by the overlaid sticks on the convergence map. In the right panel of  Fig.~\ref{fig:lensing_map} we show the measured power spectrum of the convergence (red solid lines) for sources at redshifts  $z_{\rm s} = 1.0, 2.0, 3.0$ (from bottom to top), scaled by factors of 1, 10 and 100, respectively, for clarity. The convergence power from the ray-tracing simulation shows good agreement with the predictions from the Born approximation (gray dashed lines) and halofit (black dashed lines) across weak lensing scales. The blue solid lines represent the shear E-mode power spectrum, demonstrating that $C^{\rm EE}_{\ell} = C^{\kappa\kappa)}_{\ell}$ at high-$\ell$ (small scales), consistent with the theoretical predictions. For comparison, we also show the power of the shear B-mode (green) and rotation mode (cyan) for sources at $z_{\rm s} = 1.0$, which are effectively suppressed by more than four orders of magnitude relative to the E-mode.

\begin{figure*}
    \centering
    \tikzstyle{decision} = [diamond, draw, fill=blue!20, 
text width=5.5em, text badly centered, inner sep=0pt]
\tikzstyle{block} = [rectangle, draw, fill=blue!20, 
text width=10em, text centered, rounded corners, minimum height=2em]
\tikzstyle{line} = [draw, -latex']
\tikzstyle{cloud} = [draw, ellipse,fill=red!20,
minimum height=4em]
\tikzstyle{decisionn} = [diamond, draw, fill=blue!20, 
text width=4em, text badly centered, node distance=4cm, inner sep=0pt]
\tikzstyle{data_block} = [rectangle, draw, fill=red!20, 
text width=6em, text centered, minimum height=2em]
\tikzstyle{temp_data_block} = [rectangle, draw, fill=blue!20, 
text width=6em, text centered, minimum height=2em]
\tikzstyle{final_data_block} = [rectangle, draw, fill=green!20, 
text width=6em, text centered, minimum height=4em]
\tikzstyle{module_block} = [ellipse, draw, fill=blue!20, 
text width=6em, text centered, minimum height=2em]
\tikzstyle{module_block_big} = [ellipse, draw, fill=blue!20, 
text width=8em, text centered, minimum height=2em]
\centering
\begin{tikzpicture}[node distance = 2cm, auto]
    \node[data_block] (galaxy_cat) {Extragalactic objects, SEDs};
    \node[data_block, right=2.5cm of galaxy_cat] (star_cat) {Stellar objects, SEDs};
    \node[module_block_big, below right=1.0cm and 0cm of galaxy_cat] (object_parsing) {Object parsing \& Astrometric modeling};
    \node[data_block, below right=0.6cm and 1.5cm of object_parsing] (obs_parameter) {Observation parameters};
    \node[data_block, below left=0.6cm and 0cm of object_parsing] (focal_def) {Focal plane layout};

    \node[temp_data_block, right=1cm of object_parsing] (obj_list) {MockObject list};
    \node[module_block, right=1cm of obj_list] (obj_model) {Object modeling};
    \node[module_block, above=0.6cm of obj_model] (psf_model) {PSF\&distortion modeling};
    \node[data_block, above=0.6cm of psf_model] (psf_sample) {PSF samples};
    \node[module_block, above=0.6cm of psf_sample] (optical_sim) {Optical Simulation};

    \node[decisionn, below of=obj_model, node distance=2.5cm] (decide) {Spetrum chip?};
    \node[module_block, below=0.5cm of decide] (disperse) {Spectral dispersion};
    \node[module_block, below=1.5cm of obs_parameter] (rendering) {Rendering};
    \node[module_block, left=1cm of rendering] (noise) {Noise\&readout modeling};
    \node[final_data_block, left=1cm of noise] (products) {Simulation Products};

    \path [line] (optical_sim) -- (psf_sample);
    \path [line] (psf_sample) -- (psf_model);
    \path [line] (psf_model) -- (obj_model);
    
    \path [line] (galaxy_cat) -| (object_parsing);
    \path [line] (star_cat) -| (object_parsing);
    \path [line] (obs_parameter) -| (object_parsing);
    \path [line] (focal_def) -| (object_parsing);
    \path [line] (object_parsing) -- (obj_list);
    \path [line] (obj_list) -- (obj_model);

    \path [line] (obj_model) -- (decide);
    \path [line] (decide) -- node [near start] {yes} (disperse);
    \path [line] (decide) -| node [near start] {no} (rendering);
    \path [line] (disperse) -- (rendering);
    \path [line] (obs_parameter) -- (rendering);
    \path [line] (rendering) -- (noise);
    \path [line] (noise) -- (products);
\end{tikzpicture}
    \caption{High-level overview of simulating a single CSST exposure.}
    \label{fig:sim_work_flow}
\end{figure*}
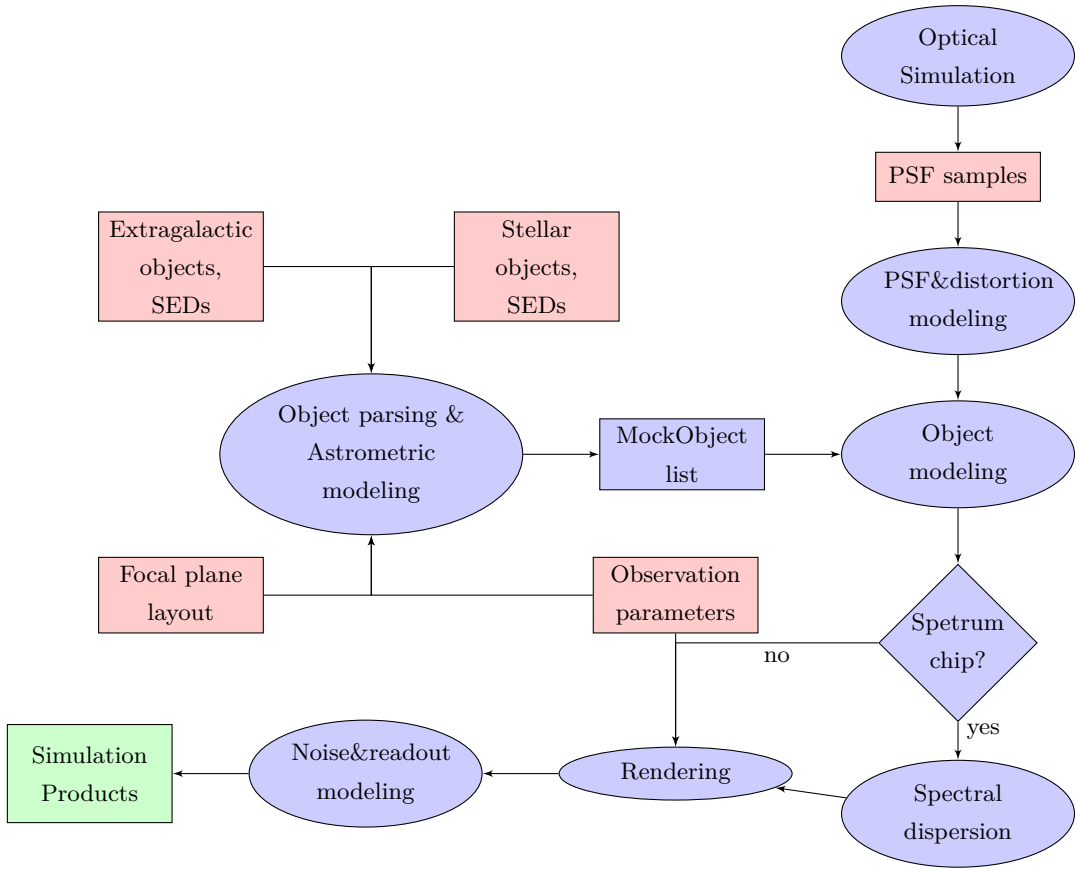

\section{The Simulation Workflow}
\label{sect:workflow}
 Fig.~\ref{fig:sim_work_flow} provides a schematic overview of the CSST simulation workflow. The process starts by extracting the parameters for an individual observation from a reference survey table prepared by the Survey Strategy Working Group. The simulation proceeds on a per-chip basis, where the sky footprint of each chip is determined and used to select relevant science objects and their associated SEDs from the True Universe catalogs. The selected objects are then parsed into a list of \texttt{MockObject} instances, which act as the primary input interface for the simulation suite.

As the list is iterated, each \texttt{MockObject} is modeled according to its specific type and properties (both intrinsic and observational). The PSFs and field distortion models are derived using samples from a dedicated simulation of the optical system and payload platform.  The model of each object is convolved with the corresponding PSF, and its shape and image position are modified based on the distortion model. If the current chip being simulated has a grism, as defined by the focal plane configuration, the object model additionally undergoes a spectral dispersion step. After completing all the above modeling processes, the final object model is rendered onto the chip image.

Finally, various observational noise components and the detector readout process are simulated, and the resulting data products are written to disk.

\subsection{Multi-band Imaging Observation}
\label{subsect:photometric_obs}

Depending on the current configuration of the CSST focal plane ( Fig.~\ref{fig:focal_plane_layout}), a photometric imaging simulation step would be performed for chips 6, 7, 8, 9, 11, 12, 13, 14, 15, 16, 17, 18, 19, 20, 22, 23, 24, and 25. 

After reading in the  parameters on observation and True Universe data, a list of \texttt{MockObject} instances is generated. This list includes all objects that are either within or very close to the sky footprint covered by the detector that is being simulated. Each instance in the list represents an object and contains detailed attributes corresponding to both its intrinsic and observed quantities. The observed quantities---such as the observed SED, apparent magnitudes, Milky Way extinction, and sky position at the observing epoch---are computed using the simulation suite's utilities, including tools for the extinction correction of the Milky Way , synthetic photometry, and astrometry.

Next, we iterate through this list, modeling  objects one by one and transforming it into a Galsim \texttt{GSObject} according to its type and corresponding parameters. The image position of each object is calculated by first projecting its sky coordinates via WCS, and then shifted by the field distortion module. A PSF model is then retrieved at the image position by interpolating the neighboring samples. This PSF model is quasi-chromatic, sampled at four specific wavelength points within the corresponding bandpass. As a result, each filter is divided into four sub-bandpasses. object is rendered into four separate image stamps by convolving its \texttt{GSObject} with the PSFs of all sub-bandpasses. These image stamps are subsequently added to the final chip image.

Once all objects have been rendered, additional processing steps are applied to simulate the effects of detector  on the chip image. These steps include the introduction of various sources of noise, such as readout noise, sky background, and cosmic rays, to create a more realistic simulation. Additionally, these procedures emulate other detector-specific characteristics like pixel response non-uniformity and brighter-fatter effects. This ensures that the final image closely resembles what would be obtained from an actual observation, effectively mimicking the desired output format for downstream analysis and testing.

\subsection{Slitless Spectroscopic Observation}
\label{subsect:spect_obs}

For the simulation of slitless spectroscopy, the process remains identical to direct imaging until the light reaches the grating. Therefore, all the steps leading up to object rendering follow the same procedures described in Section~\ref{subsect:photometric_obs}. 

The next step involves performing light dispersion for all rendered objects. Based on the characteristics of the CSST slitless spectroscopy, we simulate 24 gratings (there are 2 gratings for each chip) individually and derive the spectral dispersion properties according to the position of the incident light. In this process, simulations are conducted for five spectral orders at each incident position: -2nd, -1st, 0th, 1st, and 2nd orders. Among these, the 0th and 1st orders serve as the primary operating modes of the CSST slitless spectrometer, containing most of the incident light flux. The higher-order spectra, such as the -2nd, -1st, and 2nd orders, carry minimal energy but may introduce contamination to the spectra of the working orders. 

For the spectral characterization of different orders, we adopted the descriptive methodology utilized by the aXe extraction software \citep{2009PASP..121...59K}. 
Based on the uniform dispersion characteristics of the grating, we use linear polynomials to describe the positional relationships of the spectrum and the correlation between wavelength and position \citep{zhangxin2025}. Equation~\ref{equ:sls_pos} represents the positional relationship of the grating on the image, where (x, y) represents the position of the spectrum relative to a reference point ($x_{ref}, y_{ref}$). We choose the reference point to be the position where the corresponding spectrum is directly imaged on the focal plane.

\begin{equation}
  \label{equ:sls_pos}
  y = a_0 + a_1 x
\end{equation}
Equation~\ref{equ:sls_wave}  expresses the relationship between wavelength and position, with L representing the spectrum's length at a given position.  The spectrum length is measured from the reference point, and L can be expressed as $L = \sqrt{x^2+y^2}$.
\begin{equation}
  \label{equ:sls_wave}
  \lambda = \alpha_0 + \alpha_1 L
\end{equation}
The dispersion characteristics vary depending on the position where the light is incident on the grating. Therefore, the parameters $a_0$ and $a_1$ in Equation~\ref{equ:sls_pos} and the parameters $\alpha_0$ and $\alpha_1$ in Equation~\ref{equ:sls_wave} can be modeled as polynomials dependent on the detector's incident light position. These relationships are described by Equation~\ref{equ:pos_efficiency_a0} and Equation~\ref{equ:wave_efficiency_a0}, where ($m$,$n$) represents the CCD coordinates of the incident light.
\begin{equation}
  \label{equ:pos_efficiency_a0}
  a_k = b_{k,0} + b_{k,1}m + b_{k,2}n + b_{k,3}m^2 + b_{k,4}n^2 + b_{k,5}mn  
\end{equation}

\begin{equation}
  \label{equ:wave_efficiency_a0}
  \alpha_k = \beta_{k,0} + \beta_{k,1}m + \beta_{k,2}n + \beta_{k,3}m^2 + \beta_{k,4}n^2 + \beta_{k,5}mn  
\end{equation}

We selected a 10 $\times$ 10 grid of sampling points within a single grating area. Through optical simulations, we obtained spectral trajectory characteristics for 100 distinct positions. For each spectral trajectory, eight wavelength points were uniformly selected for characterization. Using these data, we applied the aforementioned method to fit the spectral trajectories, and compared the fitting results with the input data. The positional deviations between the same wavelengths in the working diffraction orders (0th and 1st orders) on the image were no greater than 0.2 pixels~\citep{zhangxin2025}.

Furthermore, to replicate the unique design of CSST slitless spectroscopy, our simulation incorporates bidirectional dispersion and a 1° rotation around the detector center. Bidirectional dispersion is achieved by simulating the different spectral dispersion characteristics of two gratings on one detector, while the 1° rotation is simulated by defining for each detector a specialized WCS (World Coordinate System). 

\subsection{Simulation Orchestration}

\begin{figure*}
    \centering
    \tikzstyle{decision} = [diamond, draw, fill=blue!20, 
text width=5.5em, text badly centered, inner sep=0pt]
\tikzstyle{block} = [rectangle, draw, fill=blue!20, 
text width=10em, text centered, rounded corners, minimum height=2em]
\tikzstyle{line} = [draw, -latex']
\tikzstyle{cloud} = [draw, ellipse,fill=red!20,
minimum height=4em]
\tikzstyle{decisionn} = [diamond, draw, fill=blue!20, 
text width=6em, text badly centered, node distance=4cm, inner sep=0pt]
\tikzstyle{data_block} = [rectangle, draw, fill=red!20, 
text width=6em, text centered, minimum height=2em]
\tikzstyle{temp_data_block} = [rectangle, draw, fill=blue!20, 
text width=6em, text centered, minimum height=2em]
\tikzstyle{final_data_block} = [rectangle, draw, fill=green!20, 
text width=6em, text centered, minimum height=4em]
\tikzstyle{module_block} = [ellipse, draw, fill=blue!20, 
text width=6em, text centered, minimum height=2em]
\centering
\begin{tikzpicture}[node distance = 2cm, auto]

\node[data_block] (obs_config) {Observation configuration YAML};
\node[module_block, below=1.cm of obs_config] (init) {Intialization};
\node[draw, fill=blue!20, minimum width=7cm, minimum height=1.5cm, below=1cm of init] (status) {};
\node[anchor=north] at (status.north) {Status};
    \node[rectangle, draw, fill=blue!5, rounded corners, minimum size=1.cm] at (status.south west) [anchor=south west, xshift=0.1cm](chip) {Chip};
    \node[rectangle, draw, fill=blue!5, rounded corners, minimum size=1.cm, right=0cm of chip] (filter) {Filter};
    \node[rectangle, draw, fill=blue!5, rounded corners, minimum size=1.cm, right=0cm of filter] (telescope) {Telescope};
    \node[rectangle, draw, fill=blue!5, rounded corners, minimum size=1.cm, right=0cm of telescope] (pointing) {Pointing};
    \node[rectangle, draw, fill=blue!5, rounded corners, minimum size=1.cm, right=0cm of pointing] (catalog) {Catalog};

\node[data_block, right=3cm of init] (step_list) {List of (layer, obs\_param)};
\node[decision, below=1cm of step_list] (decide) {is empty?};
\node[ellipse, draw, fill=blue!20, text width=8em, text centered, minimum height=2em, below=1cm of status] (run_step) {Call next\\ layer(obs\_param)};
\node[data_block, below=1cm of run_step] (products) {Products};
\node[module_block, below=1cm of decide] (finish) {Finish};

\path [line] (obs_config) -- (init);
\path [line] (init) -- (status);
\path [line] (init) -- (step_list);
\path [line] (step_list) -- (decide);
\path [line] (status) -- (run_step);
\path [line] (decide) -- node [near start] {no} (finish);
\path [line] (decide) -- node [near start] {yes} (run_step);
\draw[->] (run_step) to[out=135, in=-135] node[midway, left] {update} (status);
\path [line] (run_step) -- node [near start] {if any} (products);

\end{tikzpicture}
    \caption{A flowchart illustrating the orchestration of the simulation workflow.}
    \label{fig:sim_steps}
\end{figure*}
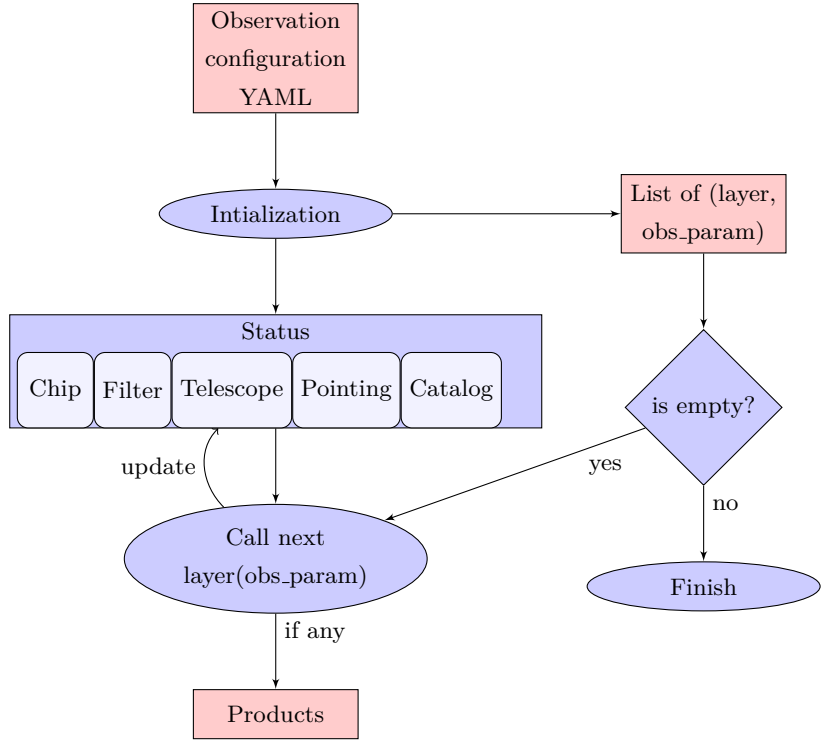

To maximize code reuse, avoid redundant development, and most importantly, accommodate the simulation requirements for various exposure modes, we structured the simulation code using "step-based" intermediate layers. Essentially, These intermediate layers encapsulate the calls to basic modules from the simulation suite, making it feasible not only to configure the parameters of individual simulation steps, but also to create different arrangements of these steps. 

The simulation layers are implemented as methods within the \texttt{SimSteps} class and are designed to run sequentially. Each layer  receives as input the current  state of the simulation, represented by a set of class instances: a \texttt{Chip}, a \texttt{Filter}, a \texttt{Telescope}, a \texttt{Pointing} and a \texttt{Catalog}.  Collectively, these objects  provide a comprehensive definition of the simulation’s context at that moment, encapsulating both the hardware configuration and observational parameters, as well as the source catalog being used. As each layer executes, it returns this state, potentially modified, which is then passed to the next layer in the sequence. This structure enables an iterative and modular transformation process throughout the simulation pipeline.

The order in which the layers are executed is configurable and can be specified via a YAML configuration file tailored for each individual observation. Additionally, an \texttt{obs\_param} dictionary is provided as input parameters, allowing for custom configurations and adjustments specific to each layer's behavior. This flexible configuration enables users to fine-tune the simulation to meet a variety of observational scenarios and requirements.

 Fig.~\ref{fig:sim_steps} presents an overview of the simulation orchestration framework. A summary of the currently implemented layers, along with their functionalities and parameters, is provided in Appendix~\ref{sect:simulation_layers}.

\section{The Simulation Suite}
\label{sect:suite}

To support the generation of realistic, pixel-level mock observations for the CSST Survey Camera, we  developed a modular simulation suite composed of specialized components. Each module is designed to model key physical and instrumental effects relevant to CSST imaging and spectroscopic observations. This section details the implementation and functionality of these core components, including PSF modeling, field distortion, object rendering, noise and detector effects, and astrometric corrections. Together, these modules enable end-to-end simulation of the CSST data acquisition process,

\subsection{Point-spread Function}
This section presents an overview of how we estimate the spatially-varying PSF of the CSST within  a field of view of SCam. 

\subsubsection{System-level PSF Simulation}
The PSF represents how a focused optical imaging system responds to a point source. The PSF of a perfect optic system is essentially the square of the modulus of the Fourier transform of the aperture function. However, in practical systems, the PSF is not just limited by diffraction but also by optical aberrations.

\begin{figure*}
    \centering
    \includegraphics[width=0.9\textwidth]{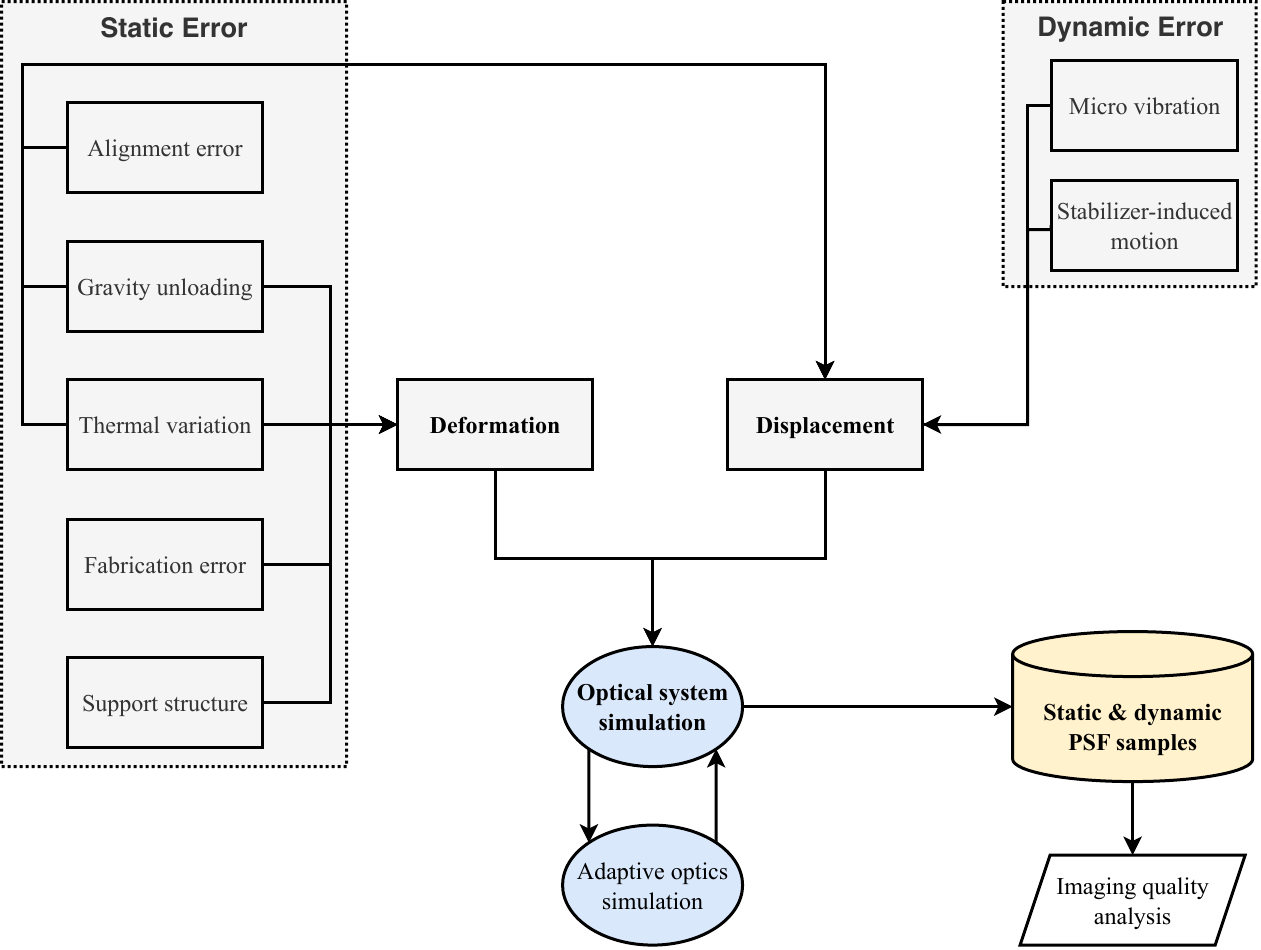}
    \caption{A schematic workflow illustrating the simulation and analysis process of the CSST optical system.}
    \label{fig:optics_sim}
\end{figure*}

The aberrations in the CSST optical system can be classified into two main categories: \textbf{static errors} and \textbf{dynamic} errors. The static errors include misalignment of optical components, gravitational unloading after launch, thermal effects in orbit$^{[4]}$ \footnotetext{[4] We assume that the thermal conditions of the CSST remain stable throughout each individual exposure.}, mirror imperfections, and deformations caused by supporting structures from both the payload and platform. The dynamic errors, on the other hand, arise from factors such as micro-vibrations and residuals of the imaging stabilization system.  Fig.~\ref{fig:optics_sim} presents a schematic workflow illustrating the simulation and analysis process of the CSST optical system. We use CODE-V for ray-tracing and wavefront calculations. By integrating the aforementioned factors into our optical models, the actual wavefront aberration of the system is computed. The resulting wavefront data is then used to generate PSF images through Fourier transform methods, enabling a more accurate assessment of the telescope's imaging performance.

Module-level testing is conducted to assess the accuracy of our simulation method. For example,  The mirror surface error (MSE) is quantified using a data fusion technique that decomposes it into low-frequency and mid-to-high-frequency  components. MSE denotes deviations of the mirror surface from its intended design, resulting in degradation of the optical system’s image quality. The simulation results for the mid-to-high-frequency components show good agreement with the test data, and the vector height distribution follows a normal distribution. Additionally, the logarithmic curve of the sagittal height distribution across the full frequency domain exhibits a gradual decreasing trend. The wavefront value of the optical-mechanical-thermal integrated system is computed through simulation and subsequently validated by temperature testing. The root mean square (RMS) values of MSE obtained from simulation and experimental measurements are $0.0603$ and $0.0596$, respectively, with a relative error of less than $1.2\%$. This confirms the accuracy of the optical-mechanical-thermal integrated simulation method. For more details, we refer the reader to Ban et al.(2025, under review).

For each CSST filter, we sample the PSF across four sub-bands within a given wide band(see  Fig.~\ref{fig:throughputs}) to gain a more comprehensive understanding of chromatic effects. In the optical system simulation, the filter structure is included with its associated surface figure error. In parallel, image quality degradation due to detector tilt and defocus is quantitatively assessed by adjusting the positional parameters of the focal plane. For further details, we refer the reader to Ban et al. (2025, under review). This allows us to analyze how variations in wavelength influence the PSF, including changes in diffraction patterns, optical aberrations, and overall image quality. By evaluating the PSF across multiple wavelengths, we can assess the impact of chromatic aberration and ensure accurate image reconstruction and calibration for astronomical observations.

\subsubsection{The interpolation calculation of the PSF for the main survey }
To accurately model the optical PSF for each object in the CSST imaging simulation, we use the CSST optical imaging system to construct the optical PSF dataset at $30\times 30$ positions within $\sim 11' \times 11'$ for each CCD. Each PSF is sampled at four wavelength position spanning the current bandpass, with a resolution of $512 \times 512$ pixels and a pixel size of 2.5 $\mu$m per pixel (corresponding to $4\times$over-sampling of the CCD). To balance memory usage and computational precision requirements, we then downsample the $4\times$over-sampled PSF to a $2\times$over-sampled version before importing it into the imaging simulation. For a given source we assign the PSF by interpolating from $30\times 30$ dateset, which is then convolved with the object to produce an optical image on the focal plane. To balance interpolation accuracy and computational efficiency, we use the  Inverse Distance Weighting (IDW) interpolation method in the current image simulation. The IDW interpolation method assumes that the PSF at any position can be interpolated from the surrounding $N$ neighboring PSF samples \citep{2013A&A...549A...1G}, with the weights given by:
\begin{equation}
    w_{i} = \frac{D^{-\beta}(x_i, x_0)}{\sum_i^N D^{-\beta}(x_i, x_0)}
\end{equation}
where $D(x_i, x_0)$ is the distance between the target ($x_0$) and the $i$-th neighboring ($x_i$) PSF sample, and $\beta$ is the power index of IDW method. 

\begin{figure*}
    \centering
    \includegraphics[width=0.85\textwidth]{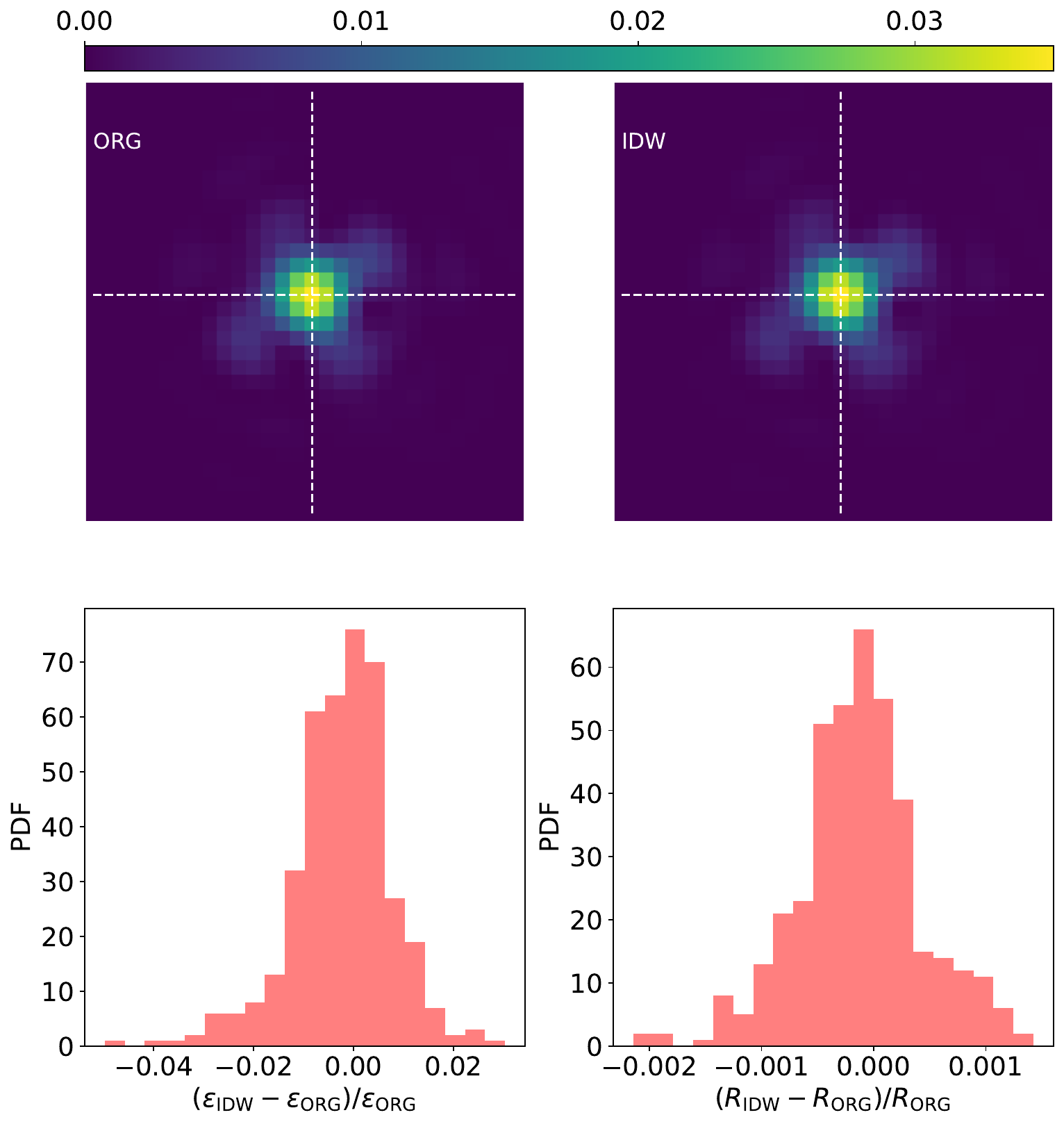} 
    \caption{Upper panels: The results of PSF interpolation. The left shows an example of the original PSF samples and the right is the PSF from IDW interpolation. Lower panels: The PDF of relative bias in ellipticity (left) and PSF size (right) for a single CCD.}
    \label{fig:psf_interpolation}
\end{figure*}

To validate the IDW interpolation method for CSST imaging simulation, a set of positions on a given CCD was selected and PSFs at these locations were generated using the IDW scheme. These interpolated PSFs were then directly compared with those produced from CSST optical system. For convenience, the validation positions were independently sampled on the focal plane following a $20 \times 20$ grid. At these positions, we compare the distribution of PSFs generated directly from CSST optical imaging system with those reconstructed through interpolation from the $30 \times 30$ reference dataset.  Fig.~\ref{fig:psf_interpolation} displays the results of PSF interpolation with $(\beta, N) = (2, 4)$. The upper panels present an example of one single PSF image from the $20 \times 20$ dataset on the left and the middle is the IDW result calculated from the $30 \times 30$ dataset at the same position. The residual image (right), $\rm (IDW-ORG)/ORG$, demonstrates that the IDW method can produce PSF images that are similar to those from CSST optical system. The lower panels are the relative bias in ellipticity (left, $\varepsilon = (Q_{11}-Q_{22}+\mathrm{i}2Q_{12})/(Q_{11}+Q_{22})$, where $Q_{ij}$ is the second brightness moments of PSF) and PSF size (right, $R = Q_{11}+Q_{22}$)  of the $20 \times 20$ positions for a single CCD. The results indicate that the ellipticity of the interpolated PSF is statistically consistent with the PSF datasets and that the relative bias in the PSF size is less than $1\%$, demonstrating that the IDW method can preserve overall image quality at a level comparable to that required for supporting data processing and scientific analysis.

\subsection{Field-distortion}\label{sect:field_distortion}
The 30$\times$30 PSF datasets as described above also incorporate the sampling coordinates $(u, v)$ on the pupil plane and the corresponding pixel coordinates $(x, y)$ on the focal plane. The pixel coordinates $(x, y)$, which contain the field distortion effect caused by the wavefront aberration, are different from the ideal pixel coordinates $(x_{0}, y_{0})$. As a result, the datasets can also be used to construct a full-view distortion model for the entire focal plane. Based on the definition of the World Coordinate System \citep[WCS;][]{2002A&A...395.1077C,2002A&A...395.1061G}, the relationship between $(x_{0}, y_{0})$ and $(u, v)$ satisfies 
\begin{equation}
\left( \begin{array}{cc} x_{0} \\
y_{0} \end{array} \right) =  
 \frac{1}{\Delta}
\left( \begin{array}{cc} \mathrm{CD2\_1} & - \mathrm{CD1\_1} \\
- \mathrm{CD2\_2} & \mathrm{CD1\_2} \end{array} \right) 
\left( \begin{array}{cc} u \\
v \end{array} \right) 
 + \left( \begin{array}{cc} \mathrm{CRPIX1} \\
\mathrm{CRPIX2} \end{array} \right),
\end{equation}
where the $\mathrm{CDi\_j}$ keywords encode the rotation and scaling in image header, and $\mathrm{CRPIX1}$ and $\mathrm{CRPIX2}$ represent the origin of the pixel coordinates along the $x-$ and $y-$axes, respectively. In addition, $\Delta$ is defined as 
\begin{equation}
\Delta = \mathrm{CD1\_2}\times \mathrm{CD2\_1} - \mathrm{CD1\_1} \times \mathrm{CD2\_2}. \nonumber
\end{equation}

The field distortion model is the mathematical mapping between $(x_{0}, y_{0})$ and $(x, y)$, which can be expressed as 
\begin{equation}
\label{eq:fdmodel1}
x = f_{x}(x_{0}, \, y_{0}); \quad y = f_{y}(x_{0}, \, y_{0}).
\end{equation}
Here we use the two-dimensional spline interpolation with an order of $n$ to approximate $f_{x}$ and $f_{y}$ throughout the focal plane, and define the residuals as $\Delta{x}=x-f_{x}$ and $\Delta{y}=y-f_{y}$.
The imperfect alignment between CCDs can result in discontinuity in the field distortion, which may not be accurately modeled by this global interpolation. Therefore, we further interpolate the residuals $\Delta{x}$ and $\Delta{y}$ on each CCD and add them back to equation~(\ref{eq:fdmodel1}) as extra correction terms. The final field distortion model can be written as 
\begin{equation}
\label{eq:fdmodel2}
\begin{split}
& x = f_{x}(x_{0}, \, y_{0}) + f_{\Delta{x}}(x_{0}, \, y_{0}); \\
& y = f_{y}(x_{0}, \, y_{0}) + f_{\Delta{y}}(x_{0}, \, y_{0}),
\end{split}
\end{equation}
where $f_{\Delta{x}}(x_{0}, \, y_{0})$ and $f_{\Delta{x}}(x_{0}, \, y_{0})$ are the interpolation functions of the residuals. We fix $n=4$ for modeling the field distortion, and refer interested readers to Ban et al. (2025, under review) for more quantitative discussions. The field distortion is found to increase from the center to the edge of the focal plane with a maximum pixel offset exceeding 200\,pixels.

In addition to the position offset, field distortion can also alter the shape of a source. Following the method presented in  \cite{2019ApJ...875...48Z}, the two ellipticity components induced by field distortion can be expressed as 
\begin{equation}
\begin{split}
& e_{1} = +(\partial_{y_{0}}y - \partial_{x_{0}}x)/(\partial_{y_{0}}y + \partial_{x_{0}}x), \\
& e_{2} = -(\partial_{x_{0}}y + \partial_{y_{0}}x)/(\partial_{y_{0}}y + \partial_{x_{0}}x).
\end{split}
\label{eq:ellmodel}
\end{equation}
Substituting equation~(\ref{eq:fdmodel2}) into equation~(\ref{eq:ellmodel}), we can calculate the ellipticity at any position on the focal plane.

\subsection{Modeling and Rendering of Objects}\label{sect:object_modeling}
In this section, we describes how galaxies, stars, and quasars are modeled and rendered in the CSST simulation.
\subsubsection{Modeling of Galaxies}
We adopted a combination of two S\'{e}rsic profiles for modeling the bulge and disk components of galaxies:
\begin{equation}
    I(r) \propto \exp(-{(r/r_e)^{1/n}})
\end{equation}
where $r_e$ is the half light radius and $n$ is the S\'{e}rsic index. In the previous cycles of simulations, namely cycle-1 to cycle-9, we have used $n=4$ for the buldge component and an exponential model with $n=1$ for the disk component. Both components are sheared to match the intrinsic ellipticity of galaxies. Then the two components are summed via:
\begin{equation}
    f_{\text{galaxy}} = (1 - b_{\text{frac}}) \times f_{\text{disk}} + b_{\text{frac}} \times f_{\text{bulge}}
\end{equation}
where $b_{\text{frac}}$ is the bulge to disk fraction.

To incorporate the deformation caused by weak lensing, parameterized by shear $(\gamma_1, \gamma_2)$, and convergence $\kappa$, The resulting profile is sheared by the reduced shear $(g_1, g_2)$, and then magnified by magnification factor $\mu$, where
\begin{equation}
    g_i = \frac{\gamma_i}{1 - \kappa}
\end{equation}
and
\begin{equation}
    \mu = \frac{1}{(1 - \kappa)^2 - (\gamma_1^2 + \gamma_2^2)}
    \label{eq:mu}
\end{equation}
When the field distortion model is enabled, additional reduced shear components $(g_{\text{fd}, 1}, g_{\text{fd}, 2})$ are added to the corresponding weak lensing reduced shear components.

The sheared profile is convolved with the respective PSF model for each of the four sub-bandpasses. The four resulting profiles are then summed to produce the final profile for the CSST band. We set the \texttt{folding-threshold} (a \textbf{GalSim} parameter that controls the maximum allowable real-space folding caused by the periodic nature of FFTs \citep{Rowe2015}) to $5e^{-8}$ for bright (with magnitude $\leq 15$) or large (with size $\geq 3$ arcsec) galaxies. For all other cases, we use  a \textbf{GalSim} default value of $5e^{-3}$.

\subsubsection{Modeling of Stars and quasars}
Stars and quasars, being typical point sources, are modeled as simple $\delta$-functions in each of the four sub-bandpasses. It is important to note that quasars are often located near the centers of their host galaxies. In our approach, we choose to model them separately, meaning that a quasar and its corresponding host galaxy are rendered and recorded as distinct objects.

Since the profiles are $\delta$-functions, we directly render the interpolated PSF onto the image for each sub-bandpass. For very bright stars and quasars (objects that are at least 1 magnitude brighter than the saturation magnitude of point sources), we extrapolated the PSF model to a larger size of $2048 \times 2048$ at its original sampling resolution to account for the photon flux in the outer regions of the stars before rendering. This approach generates a larger postage stamp for bright objects, effectively eliminating sharp transitions at the edges. Similar to the case for galaxies, we set the \texttt{folding-threshold} to $5e^{-8}$ for bright (with magnitude $\leq 15$) ones and $5e^{-3}$ otherwise.

\subsubsection{Calculation of Fluxes}
Since our PSF is modeled at four reference wavelengths within each CSST filter, the system response of each filter is divided into four sub-bandpasses. The total number of photons to be rendered for each object in each sub-bandpass is calculated via synthetic photometry. The SEDs from the truth catalog are required to be standardized to cover the wavelength range of 2000 \r{A} to 11000 \r{A} with a resolution of 0.5 \r{A}, and in units of $\text{photons}/s/m^2/\text{\r{A}}$. It is important to note that the SED used here is in the observed frame, meaning it already accounts for any potential redshift, reddening, and Milky Way extinction effects (Appendix~\ref{sect:milky_way_extinction} described in detail how we model the Milky Way extinction for galaxies). The SEDs for distant galaxies are also scaled by lensing magnification effect: 
\begin{equation}
    S(\lambda) = S_{\text{unlensed}}(\lambda)\times \mu
\end{equation}
where $S(\lambda)$ is the SED, and $\mu$ is the magnification defined in equation~\ref{eq:mu}. Then for the $i$-th sub-bandpass of filter $\alpha$, the photon fluxes is integrated using the following formula
\begin{equation}
    F_{\alpha, i} = \frac{\int S(\lambda)T_{\alpha, i}(\lambda)d\lambda}{\int T_{\alpha, i}d\lambda}
\end{equation}
where $T_{\alpha, i}$ is the transmission function over the sub-bandpass. The number of photons to be rendered is then $F_{\alpha, i} \times t_{exp} \times A_{aper}$, where $t_{exp}$ is the exposure time in seconds and $A_{aper}$ is the area of CSST aperture in $m^2$. In parallel, the AB magnitude in a filter ($\alpha$) can be calculated and recorded in the output catalog by integrating the product of the SED and the filter transmission curve over the full wavelength range of the filter. 

For spectroscopic observations, after convolving the object's profile with PSF model, the fluxes within each sub-bandpass are further dispersed into multiple orders for rendering. Readers are directed to subsection~\ref{subsect:spect_obs} for more details.

\subsection{Astrometric Effects}
To accurately simulate the apparent line-of-sight directions of stars as observed by the CSST during its orbital operations, the positions of stellar objects are systematically transformed using the microarcsecond-precision relativistic astrometric model developed by  \cite{Klioner2003AJ}. This procedure converts stellar coordinates from the International Celestial Reference System (ICRS) at a reference epoch to their apparent directions in the Geocentric Celestial Reference System (GCRS) at the observation epoch. The main steps are as follows:

\begin{enumerate}
    \item \textbf{Proper Motion Correction:} Based on stellar proper motions and, when available, radial velocities, the positions are propagated from the catalog’s reference epoch (e.g., J2000 ) to the epoch of observation, yielding the stars’ instantaneous spatial positions with respect to the Barycentric Celestial Reference System (BCRS).

    \item \textbf{Parallax Correction:} The annual parallax effect is applied together with the precise BCRS position vector of the observer at the observation epoch, obtained from high-accuracy planetary ephemerides. This transforms the barycentric direction into a geometric direction from the geocenter while retaining its expression in BCRS coordinates.

    \item \textbf{Gravitational Light Deflection:} The observed propagation direction of the starlight is adjusted to account for relativistic deflection due to gravitational fields—primarily the dominant deflection caused by the Sun, and, to a lesser extent, by other massive bodies in the solar system.

    \item \textbf{Aberration:} The apparent directional shift caused by the CSST’s motion with respect to the solar system barycenter is corrected, yielding the final observed position in the GCRS at the observation epoch.
\end{enumerate}

This transformation workflow conforms to the IAU (2000) resolutions on astrometric reference systems and achieves microarcsecond-level accuracy while remaining computationally practical. For full theoretical background, see  \cite{Klioner2003AJ}.

\subsection{Modeling of Noises and Detector effects}
In addition to astrophysical and optical modeling, a realistic simulation of the CSST observations requires accurate treatment of detector behavior and observational noise. This section outlines the key instrumental effects included in our simulation framework, which collectively shape the final image quality and data fidelity.

\subsubsection{Stray Light}
Stray light can significantly degrade the quality of astronomical observations by introducing unwanted light into the telescope’s optical system. It may lead to a reduction in imaging sensitivity, the introduction of artifacts such as ghost images, and systematic biases in photometric and spectroscopic measurements. For CSST, we developed a semi-analytical framework to quantitatively estimate the stray light arising from both off-field and in-field sources, accounting for complex orbital and observational conditions. Our approach models two categories of stray light: the off-field sources, such as the Sun, Moon, Earthshine, and other bright solar system objects; and the in-field contributors, including the Zodiacal background and internal scattering or ghosting caused by bright stars inside the field of view (FOV).

\begin{figure}
    \centering
    \includegraphics[width=0.75\textwidth]{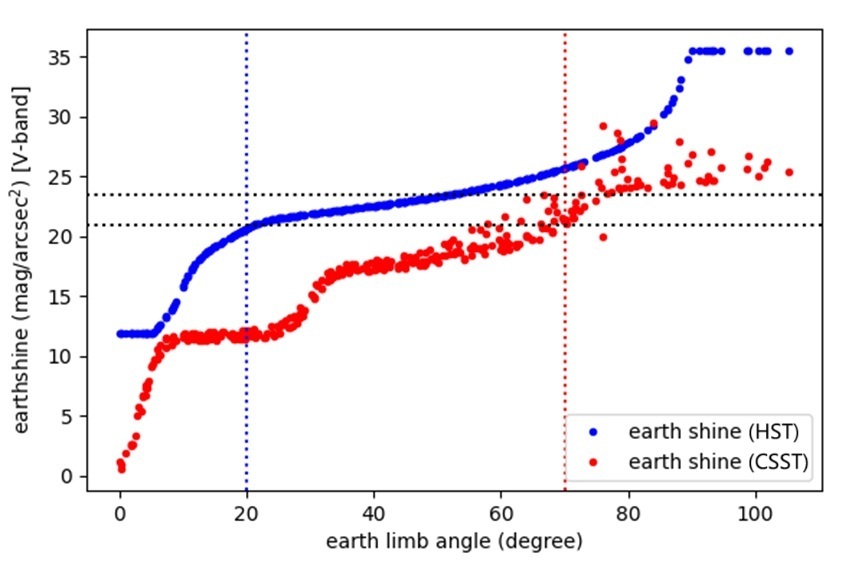}
    \caption{A comparison of Earthshine brightness in high-irradiance regions between CSST and HST shows that the Earthshine irradiance received by CSST is generally higher than that of HST across a range of observational angles.}
    \label{fig:Comparison_of_Earthshine_between_CSST_and_HST}
\end{figure}

For off-field sources, we construct a Point Source Transmittance (PST) matrix that depends on the source's zenith and azimuthal angles relative to CSST. The PST is obtained by estimating the focal plane irradiance through a hybrid approach that combines ray-tracing simulations with laboratory measurements. The irradiance from off-field point sources is subsequently calculated and corrected for obstructions caused by the Earth and the CSST aperture door. In the case of Earthshine, the Earth's surface is discretized into small segments, and both direct and scattered contributions are evaluated based on the aperture door geometry and surface reflectance properties.  Fig.~\ref{fig:Comparison_of_Earthshine_between_CSST_and_HST} presents a comparison of Earthshine brightness between CSST and HST$^{[5]}$ \footnotetext{[5] The Earthshine brightness for HST is determined using the data presented in Figure 6.2 of the STIS Instrument Handbook (https://hst-docs.stsci.edu/stisihb/chapter-6-exposure-time-calculations/6-5-detector-and-sky-backgrounds)}.

Our in-field stary light modeling includes three components:\begin{itemize}
    \item \textbf{Zodiacal light}, is modeled as a diffuse background based on observed brightness maps (\citealt{Leinert1998}).
    \item \textbf{Scattering from bright stars within the FOV}, is modeled using a linear combination of Bidirectional Reflectance Distribution Functions (BRDFs), derived from metrology-based power spectral density (PSD) measurements of each optical surface. The resulting estimates are validated against simulations performed using FRED$^{[6]}$ \textcolor{red}{\footnotetext{[6] \url{https://photonengr.com/fred} }}, an optical engineering software, showing a discrepancy of less than $7 \%$.
    \item \textbf{Ghost images}, resulting from multiple internal reflections between the detector and filter surfaces, are analytically modeled based on the measured reflectance properties of the filters and detectors.
\end{itemize}

\begin{figure}
    \centering
    \includegraphics[width=0.75\textwidth]{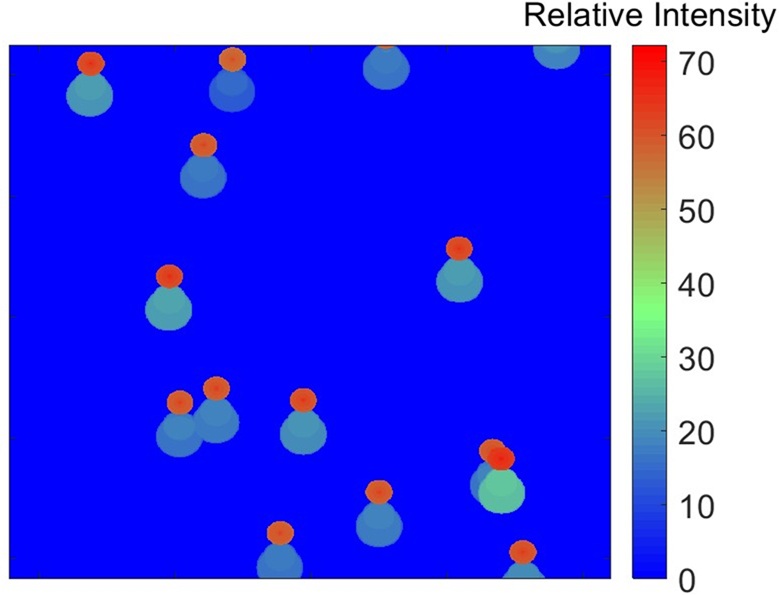}
    \caption{The simulated intensity map illustrates the scattering and ghost images originating from bright stars within the field of view. The results show that ghost image relative intensities are on the order of $10^{-6}$ or lower, while scattered light exhibits an intensity level of approximately 10$^{-4}$ at a distance of 10 pixels from the image point, decreasing to around 10$^{-6}$ at 100 pixels.}
    \label{fig:Scientific_Simulation_Image_with_In-Field_Stray_Light}
\end{figure}

 Fig.~\ref{fig:Scientific_Simulation_Image_with_In-Field_Stray_Light} illustrates effects of scattering and ghosts from bright stars within a FOV. Our framework enables rapid yet accurate estimation of stray light under dynamic survey conditions, supporting both image simulation and observation planning.

\subsubsection{Brighter-fatter effects}
\label{sect:bright_fatter}
As charge accumulation in CCDs increases, the resulting electrostatic repulsion causes stellar profiles to broaden with increasing brightness—a phenomenon known as the brighter-fatter effect \citep{2014JInst...9C3048A,2017arXiv170305823L}. This effect introduces systematic biases in galaxy shape measurements derived from star-based PSF models \citep{2023A&A...670A.118A, 2024PASP..136d5003B}, and therefore must be incorporated into realistic image simulations. 

To model this electron deflection effect, \cite{2021JAP...130p4502L} developed a grid-based Poisson solver (Poisson-CCD), whose electrostatic solutions were integrated into GalSim using a fast interpolation scheme. However, due to the high computational cost and limited adaptability of the Poisson-CCD approach for CSST-specific detector parameters, we implement an alternative phenomenological model. 

Our model approximates the brighter-fatter effect as a local smoothing operation characterized by a kernel function, whose coefficients are derived from pixel-to-pixel correlations in flat-field exposures. Since the weights of the kernel function rapidly decay to the noise level in the outer regions, we adopt a $5 \times 5$ smoothing kernel to ensure algorithmic stability. The coefficients of smoothing kernel are calculated from the covariance matrix of correlated pixels in the flat-field data. For further details, we refer the reader to Wei et al. (2025, in prep.). 

\begin{figure}
    \centering
    \includegraphics[width=0.75\textwidth]{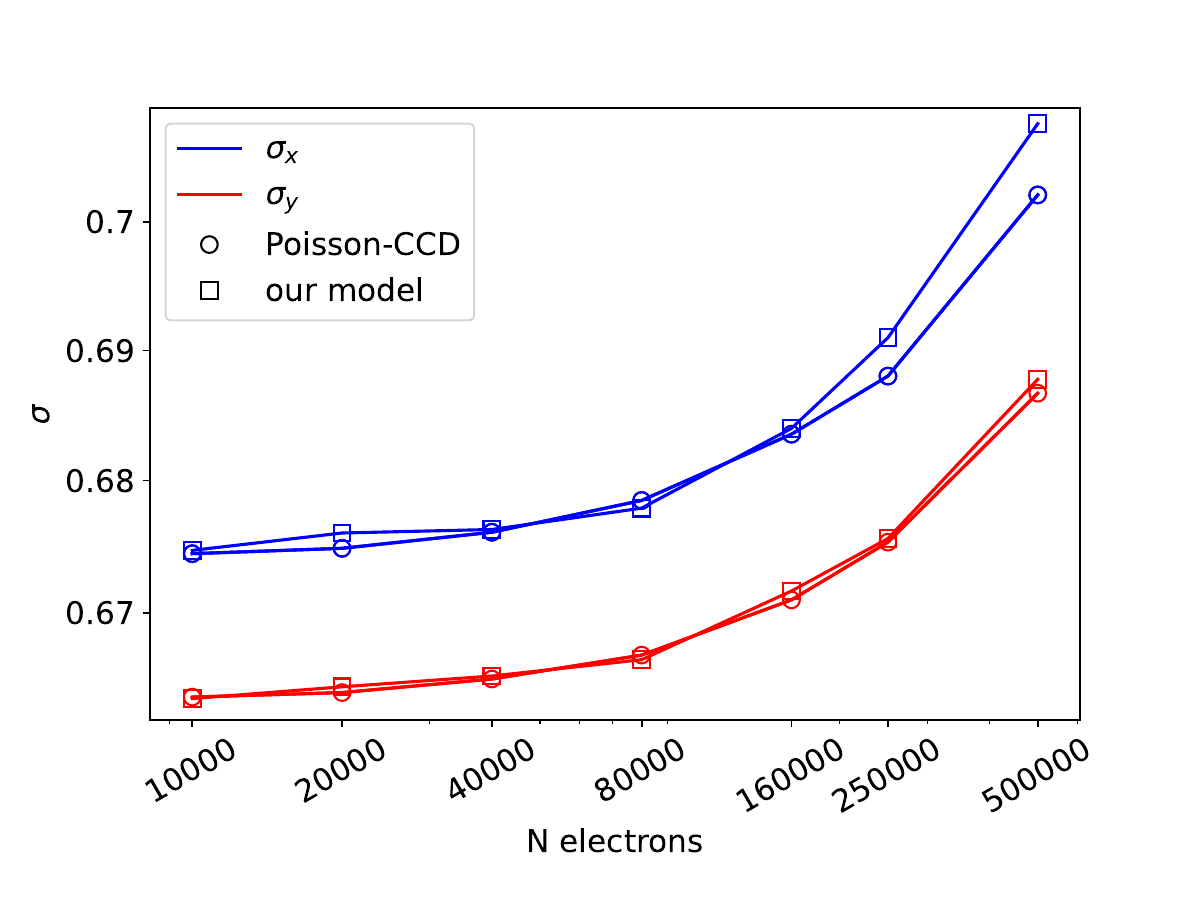}
    \caption{The comparison of our model and Poisson-CCD in simulating the brighter-fatter effect.}
    \label{fig:B-F}
\end{figure}

To validate this model, we simulate a series of stellar images with varying surface brightness using both the SiliconSensor module in Galsim and our phenomenological approach.  Fig.~\ref{fig:B-F} compares the Gaussian widths $(\sigma_x, \sigma_y)$ derived from both methods—represented by circles (Poisson-CCD) and squares (our model), respectively. The results confirm that the profile sizes increase with brightness and that our model successfully reproduces the statistical behavior of the brighter-fatter effect, demonstrating its suitability for CSST image simulations.

\begin{figure*}
    \centering
    \includegraphics[width=0.9\textwidth]{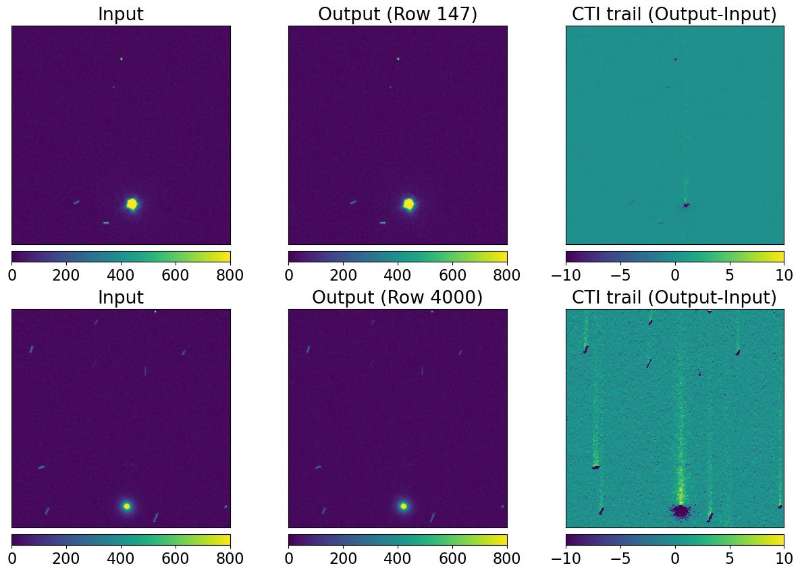}
    \caption{Input, output and CTI-induced trails for two stamps cutouts 147 rows and 4000 rows away from the channel readout.}
    \label{fig:MCTIt}
\end{figure*}

\begin{figure*}
    \centering
    \begin{subfigure}[b]{0.45\textwidth}
        \includegraphics[width=\textwidth]{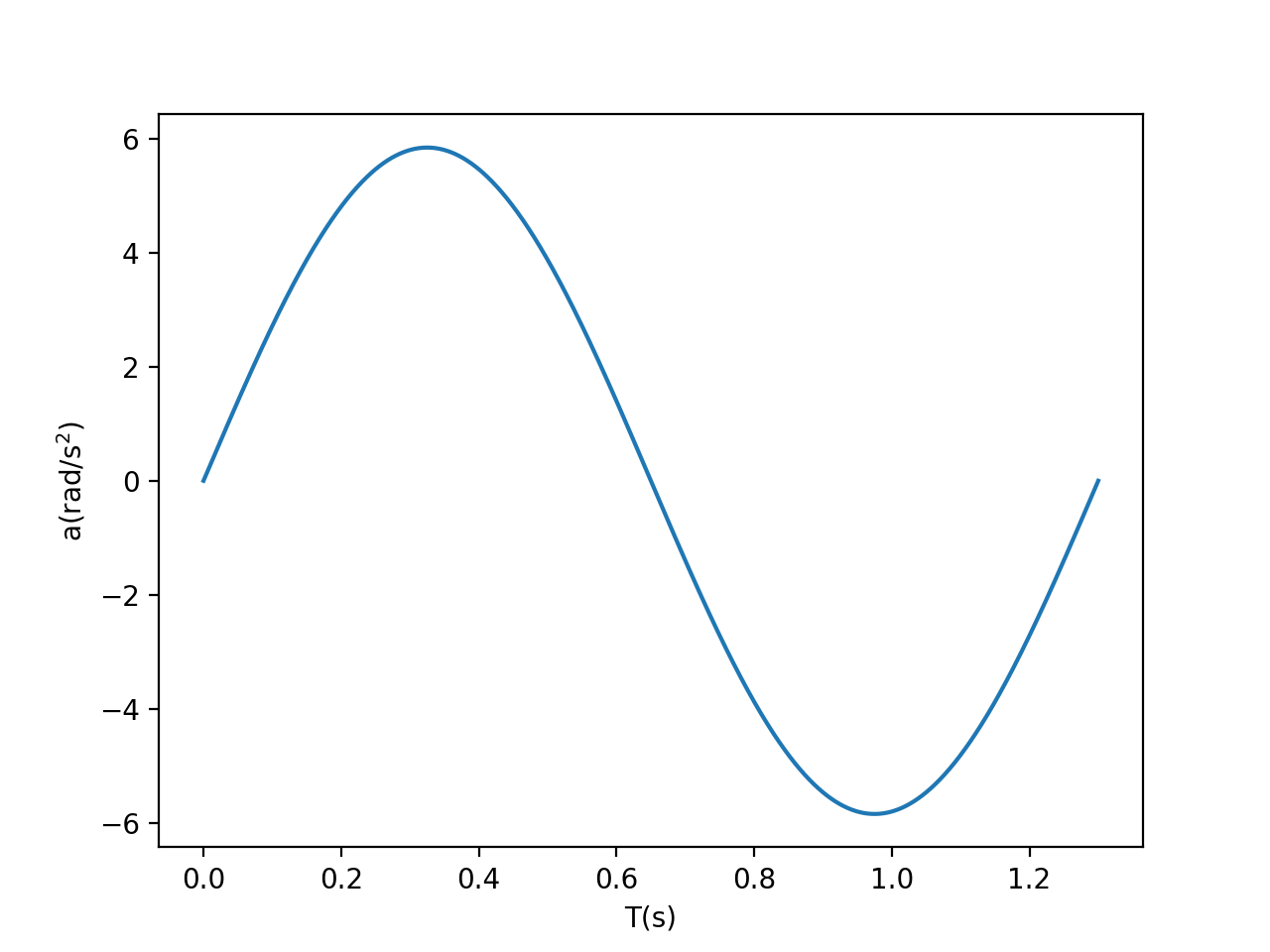}
        \caption{The angular acceleration of shutter motion.}
        \label{fig:fig_shutter1}
    \end{subfigure}
    \hspace{5pt}
    \begin{subfigure}[b]{0.45\textwidth}
        \includegraphics[width=\textwidth]{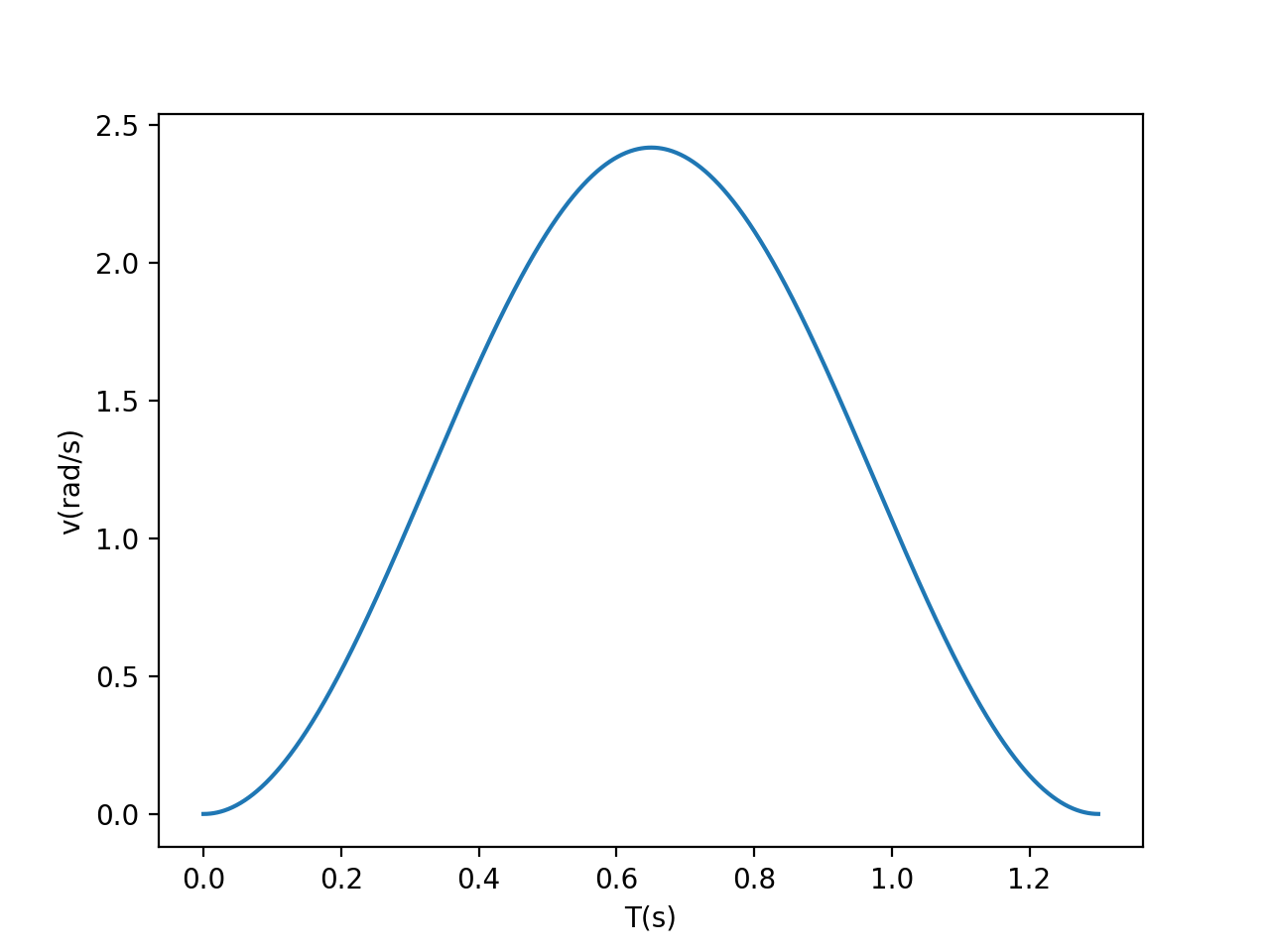}
        \caption{The angular velocity of shutter motion.}
        \label{fig:fig_shutter2}
    \end{subfigure}
    \hspace{5pt}
    \begin{subfigure}[b]{0.45\textwidth}
        \includegraphics[width=\textwidth]{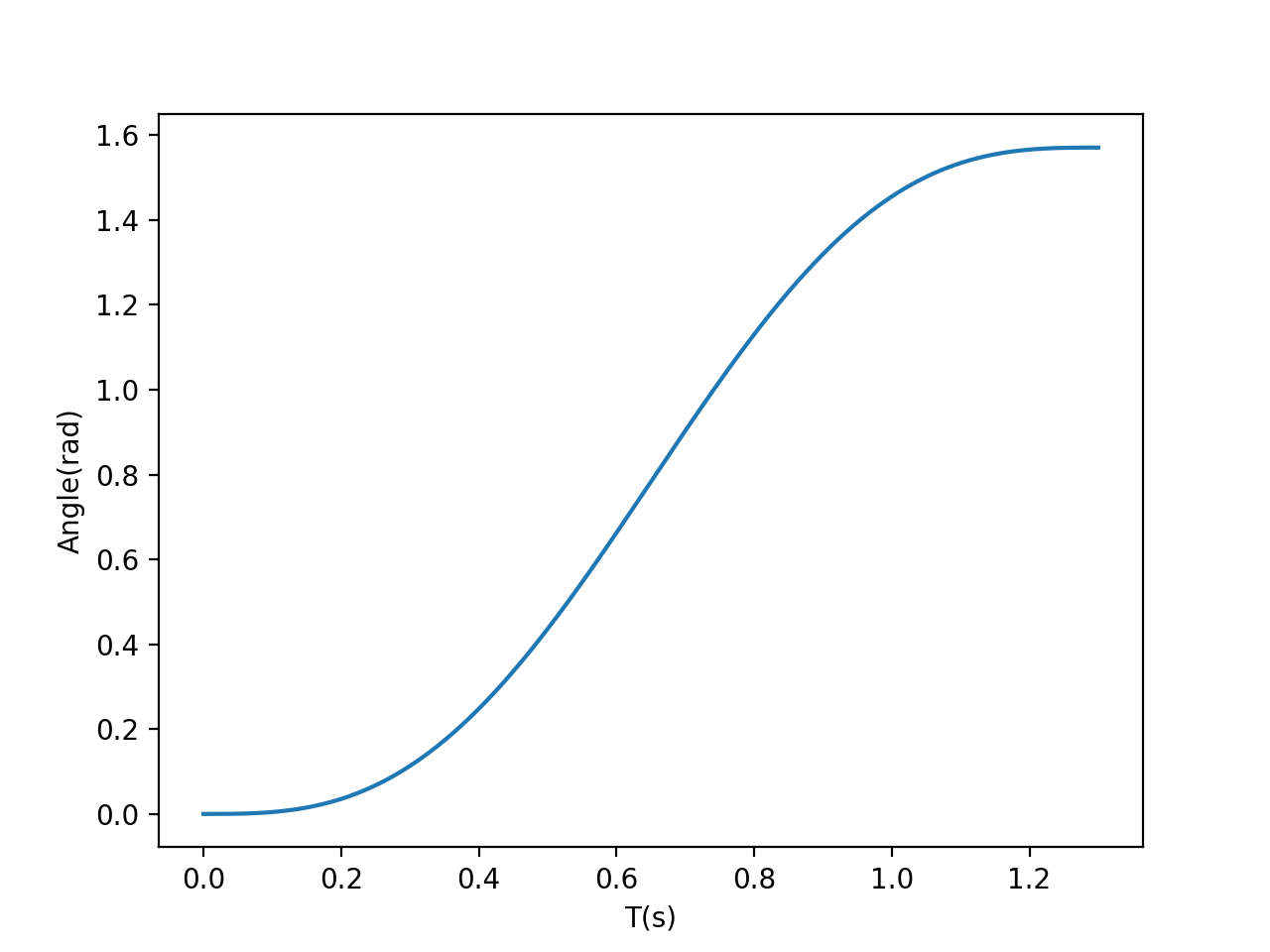}
        \caption{The profile of shutter motion.}
        \label{fig:fig_shutter3}
    \end{subfigure}
    \hspace{5pt}
    \begin{subfigure}[b]{0.45\textwidth}
        \includegraphics[width=\textwidth]{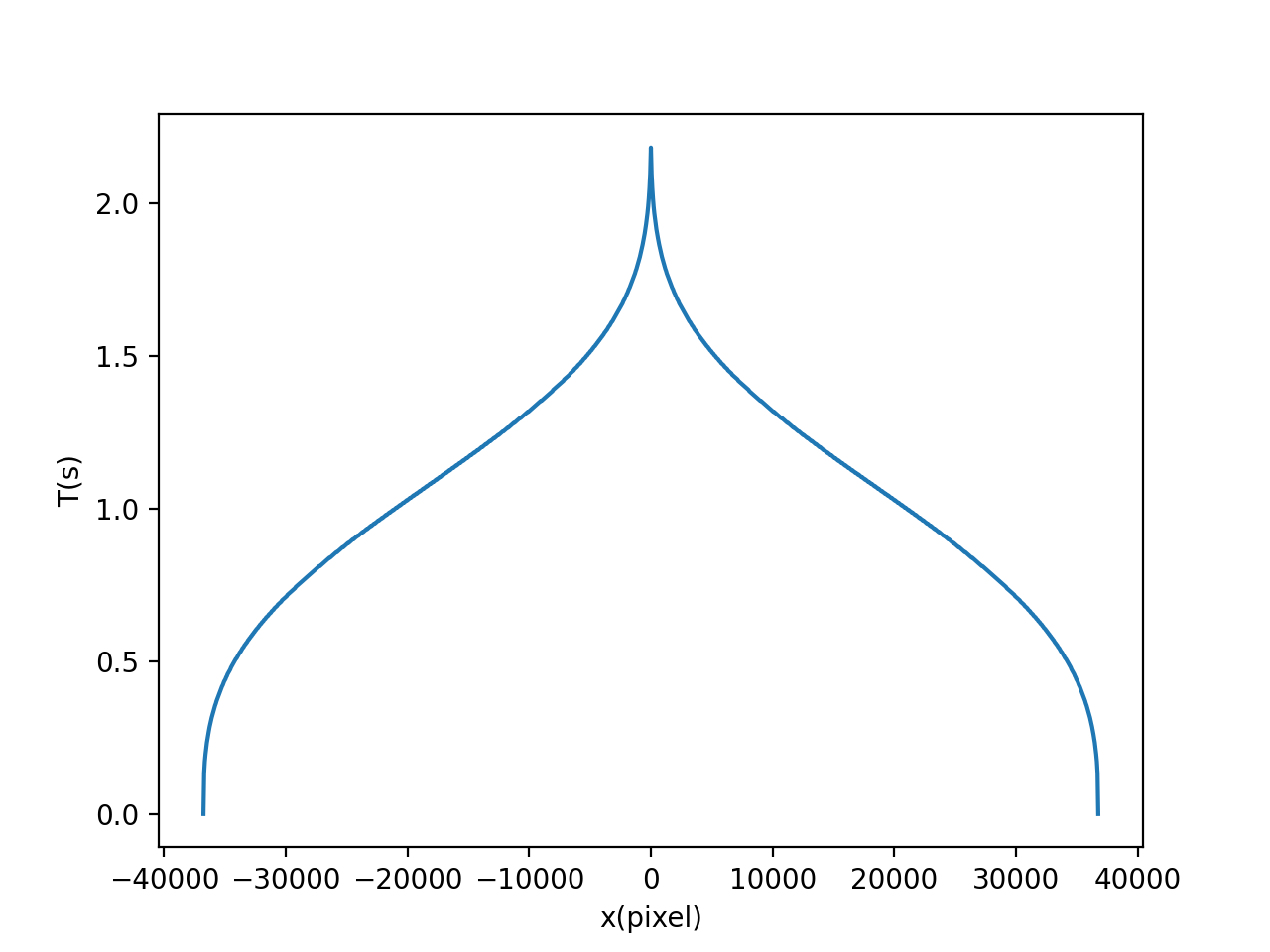}
        \caption{The net reduction in effective exposure time across pixel coordinates along the shutter operating direction due to the its opening/ closing motion.}
        \label{fig:fig_shutter4}
    \end{subfigure}
    \caption{Shutter rotation model curves (the shutter opening/closing time is 1.3s in the figure)}
    \label{fig:shutter_example}
\end{figure*}

\subsubsection{Charge transfer inefficiency effects}
\label{sect:CTI}
High energy particles from the space radiation environment can interact with CCD detectors, causing lattice displacements and creating additional vacancies. These vacancies act as electron traps during the charge transfer process—capturing a portion of the charge from one packet and releasing it into subsequent ones—resulting in image distortions. This phenomenon is known as charge transfer inefficiency (CTI).

Based on the Shockley-Read-Hall theory \citep{PhysRev.87.835,PhysRev.87.387}, several space missions, such as Gaia \citep{8185613}, HST \citep{Anderson_2010,8223496,8184054} and Euclid \citep{8200689}, have developed CTI simulation algorithms tailored to their CCD architectures. A detailed comparison of these methods can be found in \citet{8184054}. 

In this work, we adopt the modeling framework described in \citet{8223496} which assumes several discrete electron trap species, each described by its density $\rho$ and its electron release timescale $\tau$. For a given species, the number of traps within a pixel follows a Poisson distribution with a mean of $\rho$, and each trap is assigned a random watermark uniformly distributed between 0 and 1. When a charge packet with electron counts $N_\mathrm{e}$ is transferred into the pixel, the volume ratio $V$ it occupies is given by
\begin{equation}
V = [\mathrm{max}(N_\mathrm{e}-n,0)/w]^\beta
\end{equation}
where $w$ is the fullwell capacity and $n$ is the notch channel depth. Each empty trap (one that has not captured an electron since its last release) with its watermark below $V$ will capture one electron (instantaneous capture), while each filled trap with a watermark above $V$ has a probability $1-e^{-1/\tau}$ of releasing an electron. Laboratory tests on E2V CCD290-99, one of the CSST Survey Camera CCD candidates, show that a triple trap species model with $\beta\sim0.5,n\sim0$ can effectively describe its CTI effect. 

To validate this model, we apply the CTI effect to a field image with $4616\times4608$ pixels and check the output. The model parameters are set to be $\beta=0.492$, $n=0$, $w=85200\mathrm{e}$, $\tau=\{1.22,14.02,84.74\}$ and $\rho=\{0.025,0.105,0.655\}$.  Fig.~\ref{fig:MCTIt} shows two input image stamps located 147 rows and 4000 rows away from the channel readout along with their corresponding output images and their CTI-induced trails. As expected, sources farther from the readout end show longer CTI trails. We compare our output image with that of the public available code ArCTIc$^{[7]}$ \footnotetext{[7] \url{https://github.com/jkeger/arctic}}, and the results are consistent. For further details, please refer to Luo et al. (2025, in prep.).

\subsubsection{Shutter Effect}
\label{sect:shutter}
For the shutter effect, we assumed that the rotation angle of the shutter from closed to open is $0^\circ - 90^\circ$ (in fact, the maximum opening angle may vary slightly to account for manufacturing tolerances, assembly stack-up errors, and oscillation avoidance during closure sequencing) and the acceleration  of the shutter ($a$) is a sinusoidal function of time t as shown in Equation~\ref{equ:shutter}. Then one can derive the velocity ($v=\int_{0}^{t} adt$) and the rotation angle of the shutter ($A_{\rm shutter}= \int_{0}^{t} vdt$), respectively.  Fig.~\ref{fig:fig_shutter1},\ref{fig:fig_shutter2},\ref{fig:fig_shutter3} illustrates the resulting time-dependent profiles of acceleration, velocity, and rotation angle throughout the full opening cycle. The shutter opening time is set to be 1.3 seconds, which is a typical value designed for the opening and closing of the CSST shutter.

\begin{equation}
  \label{equ:shutter}
  \begin{split}
  & a = \frac{\pi^2}{t_{\rm shutter}^2} \sin\left(\frac{2\pi}{t_{\rm shutter}} t \right)
  \end{split}
\end{equation}

The focal plane layout and size of CSST are shown in  Fig.~\ref{fig:focal_plane_layout}. In the figure, the slitless spectral regions are on the left and right sides, and the shutters are located outside the slitless spectral regions I and II, respectively. The distance between the bearings of the two shutters on both sides is 735 mm. The pixel size of the focal plane is 10 $\mu m$ and the centers of the two shutters are located at the center of the focal plane, therefore, at the center of the focal plane, due to the opening and closing of the shutters, the difference between the exposure times at the center of the focal plane and  near the shutter rotation axis is twice the shutter opening/closing time.  Fig.~\ref{fig:fig_shutter4} presents the impact of shutter opening/closing on the exposure time of pixels at different positions along the shutter’s opening/closing direction,  due to the shutter effect shown in  Fig.~\ref{fig:fig_shutter1},\ref{fig:fig_shutter2},\ref{fig:fig_shutter3} (calculated for a shutter opening time of 1.3 seconds)

\begin{figure*}
    \centering
    \includegraphics[width=0.9\textwidth]{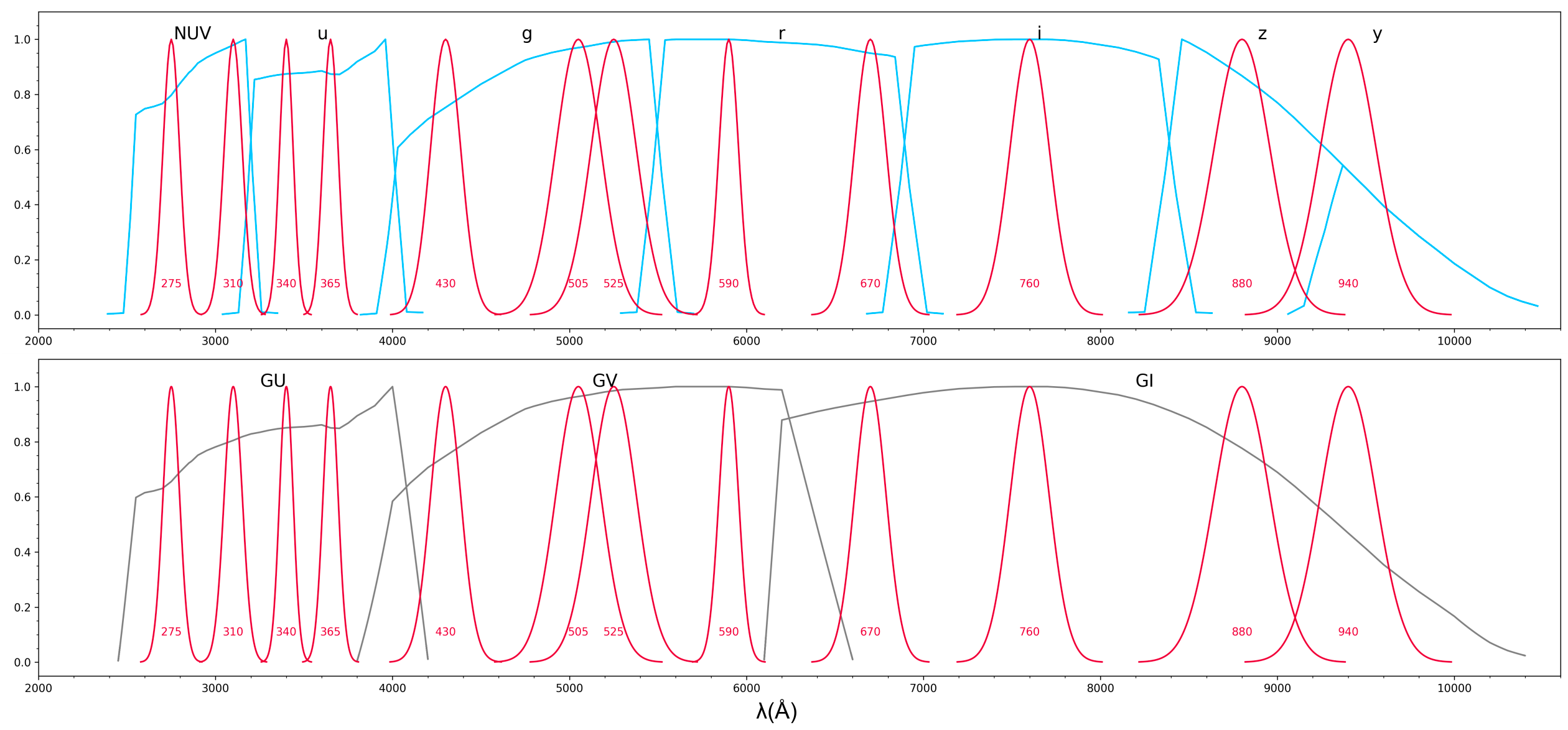}
    \caption{The 12 LEDs correspond to specific central wavelengths and wavelength ranges, which are mapped to the corresponding CSST bands. }
    \label{fig:Led_wave}
\end{figure*}

\begin{figure}
    \centering
    \includegraphics[width=0.75\textwidth]{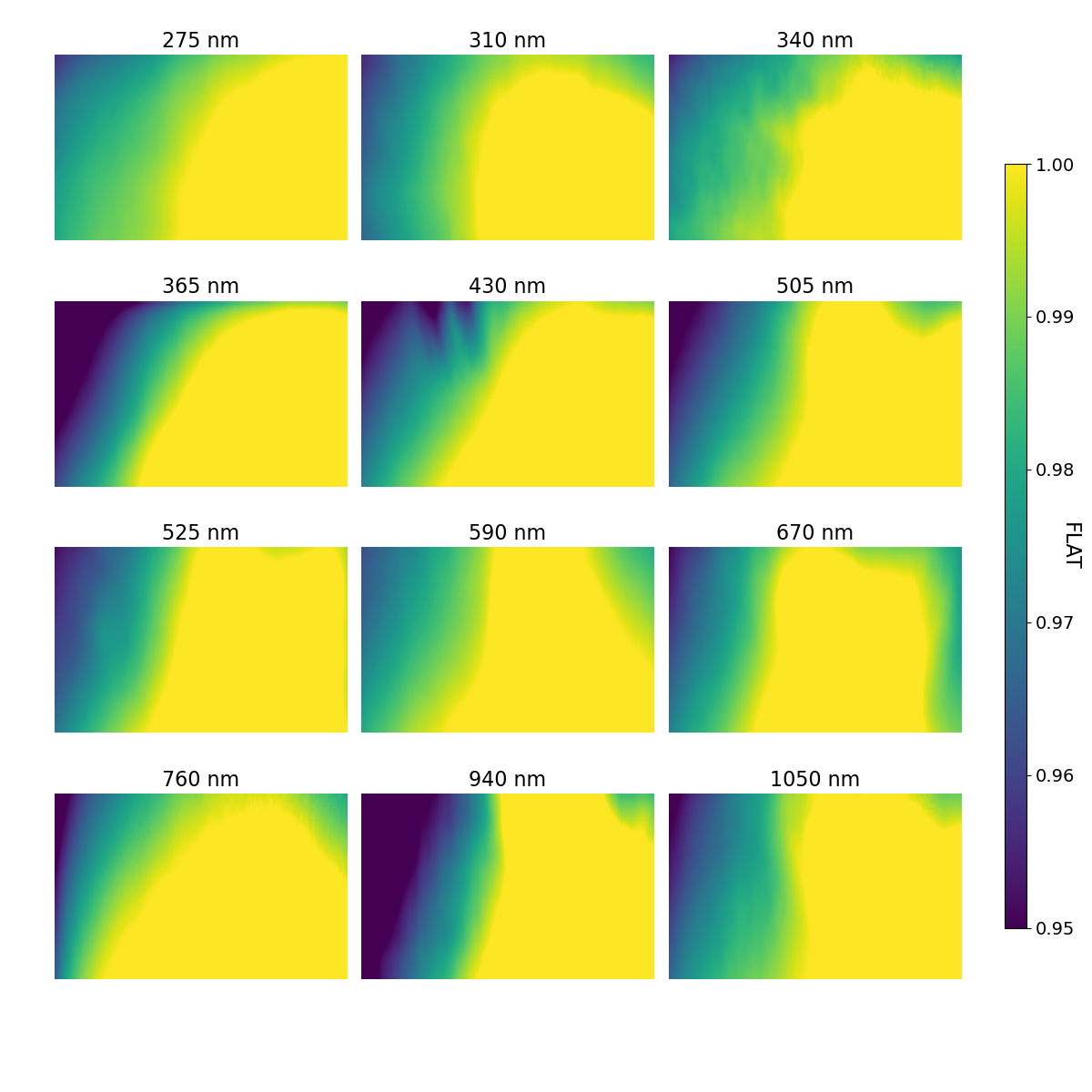}
    \caption{The flat-field images of the 12 LEDs tested in the laboratory. The central wavelengths of the LEDs correspond to those in  Fig.~\ref{fig:Led_wave}, except for the 880 nm LED, for which the flat-field image was not acquired due to experimental constraints. Instead, an additional flat-field image at 1050 nm was included. The central wavelengths of the LEDs are 275 nm, 310 nm, and 340 nm, 365 nm, 430 nm, 505 nm, 525 nm, 590 nm, 670 nm, 760 nm, 940 nm and 1050 nm respectively. }
    \label{fig:Led_flat}
\end{figure}
\subsubsection{Flat-field}
\label{sect:flat}
The flat-field model consists of two components: the external flat-field and the internal flat-field. For the external flat-field, we construct a full focal-plane flat-field vignetting model (quadratic polynomial) to characterize optical shading gradients across the detector array. This model enables the generation of vignette weight maps tailored to specific detector positions and regions, which are subsequently applied to any exposures with shutter opened for irradiance uniformity correction. The parameters of the vignetting function are generated using pseudorandom number sequences with a predefined random seed specified in the configuration file, ensuring deterministic parameter generation and reproducibility of the vignette pattern during both simulation and validation phases.

The internal flat-field accounts for response variations caused by the telescope’s internal structure, detectors, and other hardware components. In the simulation, the test results of a set of internal calibration LED light sources designed for the CSST are used as the input for constructing the internal flat-field in various wavelength bands. Within the wavelength range of the CSST survey observations, there are a total of 12 narrowband LED lamps, with their wavelength coverage shown in  Fig.~\ref{fig:Led_wave}.

The internal flat-field model is currently based on LED images acquired in a laboratory setting. The optical path in the lab replicates the CSST design, and a single detector is used to capture the irradiance, effectively mimicking the flat-field response across the full focal plane. Each LED light of different colors is used to obtain the corresponding flat-field image. These images, shown in  Fig.~\ref{fig:Led_flat} are interpolated to construct a $6 \times 5$ detector grid, forming the internal flat-field pattern for the entire focal plane.

\subsubsection{Cosmic Ray}
\label{sect:cosmicRay}
We estimate that the initial fraction of cosmic ray-impacted pixels is approximately 0.75‰. However, charge diffusion effects are projected to amplify this fraction by a factor of 2–3 in the final measurements. To realistically model their impact, we incorporate a distribution of event lengths and energies based on the statistical study by Adam Riess \citep{Riess2002}, including a subset of longer events. Specifically, 95\% of the cosmic rays have lengths randomly drawn between 0 and 20 pixels, while the remaining 5\% have lengths randomly drawn between 20 and 100 pixels.  The base-10 logarithm of the event energy is assumed to follow a Gaussian distribution with a mean of 3.3 (equivalent to 10$^{3.3}$ e$^-$) and a full width at half maximum (FWHM) of 0.6. Upon striking the detector, cosmic rays induce a certain degree of charge diffusion, leading to a spatial broadening of the cosmic ray traces. This broadening is modeled by convolving the events with a Gaussian profile of $\sigma$ = 0.2 pixels where one pixel corresponds to roughly 0.074 arcseconds.

\subsubsection{Bad Columns}
\label{sect:badCol}
We simulate bad columns in dark frames, flat-fields, and astronomical images to support early-stage testing of data processing systems. The pixel values in bad columns are set to range from [1.3$\times$Mean(background)+50e$^-$] to [2$\times$Mean(background)+150e$^-$], ensuring they are visibly detectable in all image types. Internal noise within these columns is modeled as Gaussian-distributed with a standard deviation twice that of the background noise. For each e2v 290 CCD, 1 to 5 bad columns are introduced in both the upper and lower sections. Their positions are randomly assigned within 5\% to 95\% of the total column count, and their lengths span 10\% to 70\% of the total row count. To ensure inter-chip variability, bad column patterns are generated using random seeds defined as the base seed plus the detector ID.

\subsubsection{Defective Pixel}
\label{sect:defectivePix}
Defective pixel including both dead pixels and hot pixels (with dark current > $0.1~e^-/s$), collectively accounting for $5 \times 10^{-5}$ of the total pixel population. Each type can be independently enabled or disabled. Dead pixels exhibit uniformly distributed counts between 0 and 70\% of the background mean, superimposed with Gaussian random noise ($\sigma = 5e^-$). Hot pixels are modeled using a Gamma distribution with shape parameter $2 e^-$ and scale parameter $3750e^-$. Future simulations will improve hot pixel modeling by incorporating empirical dark current distribution profiles derived from CCD characterization tests. This will lead to more accurate refinement of the model in upcoming versions. The spatial placement of defective pixels is fixed using deterministic random seeds, while variability between detectors is achieved by offsetting the seed with the detector ID.

\subsubsection{Other Effects}
\label{sect:other_effect}
The simulation includes additional detector characteristics to reflect realistic performance: Dark current is modeled at the design-specified nominal rate of $0.02 e^-/s/pixel$, while readout noise follows a Gaussian distribution with a standard deviation of $\sigma$ = $5 e^-/pixel$. The detector has a physical full-well capacity of 90,000 $e^-$ and a digital saturation level of 65,535 DN. In practice, due to the specific design of the readout circuitry, the detector almost never reaches both the physical full-well capacity and the digital saturation level simultaneously. Emulating the e2v 290 architecture, the $9\rm K\times9\rm K$ pixel array is partitioned into 16 readout channels with the following non-idealities:
\begin{itemize}
    \item Channel gains values are drawn from a uniform distribution with a mean of 1.1 and a relative variation of 1\%.
    \item Channel offsets demonstrate uniform baselines with mean = $500~e^-$ and peak-to-valley (PV) variation = $20~e^-$.
    \item Inter-channel variability is preserved through deterministic randomization using fixed seeds defined as the base seed plus the channel ID.
\end{itemize}

\section{Data Products}
\label{sect:products}
\begin{figure*}[ht]
    \centering
    \includegraphics[width=0.75\textwidth]{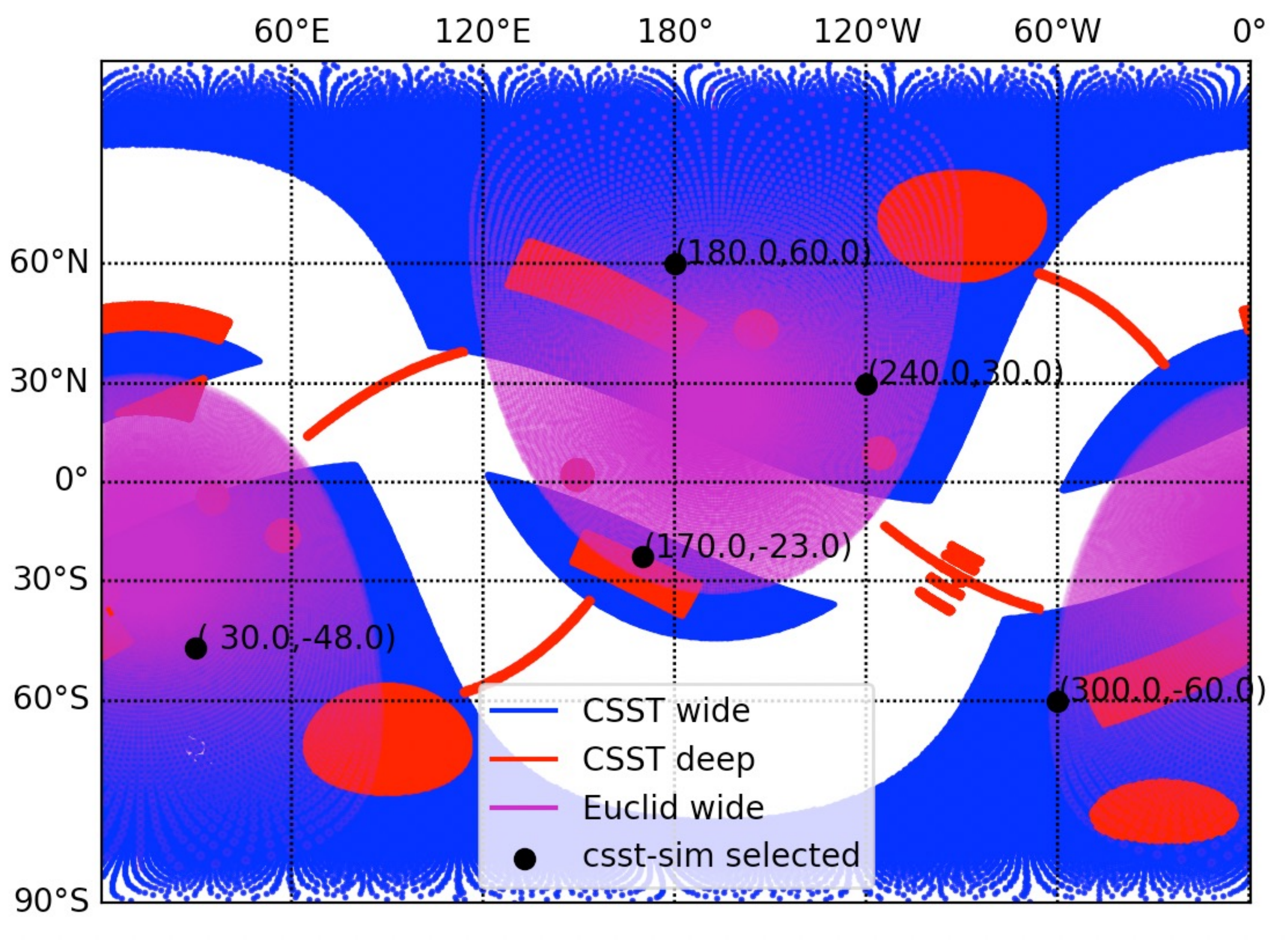}
    \caption{The distribution of the simulated field on the sky for C9 data products (black dots).}
    \label{fig:skysimulated}
\end{figure*}
\begin{figure*}[ht]
    \centering
    \includegraphics[width=0.95\textwidth]{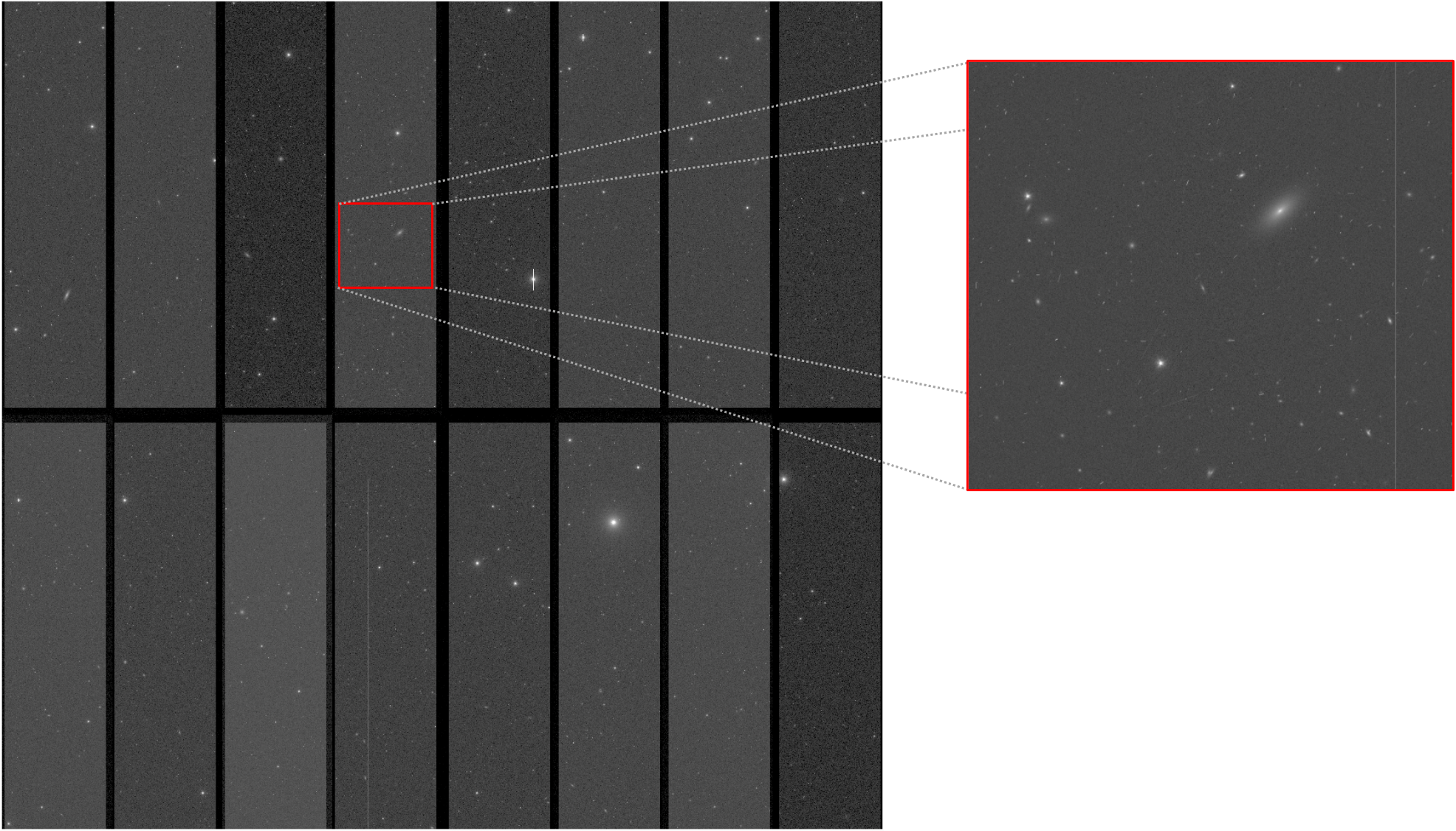}
    \caption{Log-scaled simulated image in $i$-band, with an inset showing a zoom-in region for details. Cosmic rays and some instrumental features such as bias, prescan, overscan, $2 \times 8$ readout channels are included.}
    \label{fig:csstsim_24}
\end{figure*}
\begin{figure*}[ht]
    \centering
    \includegraphics[width=0.95\textwidth]{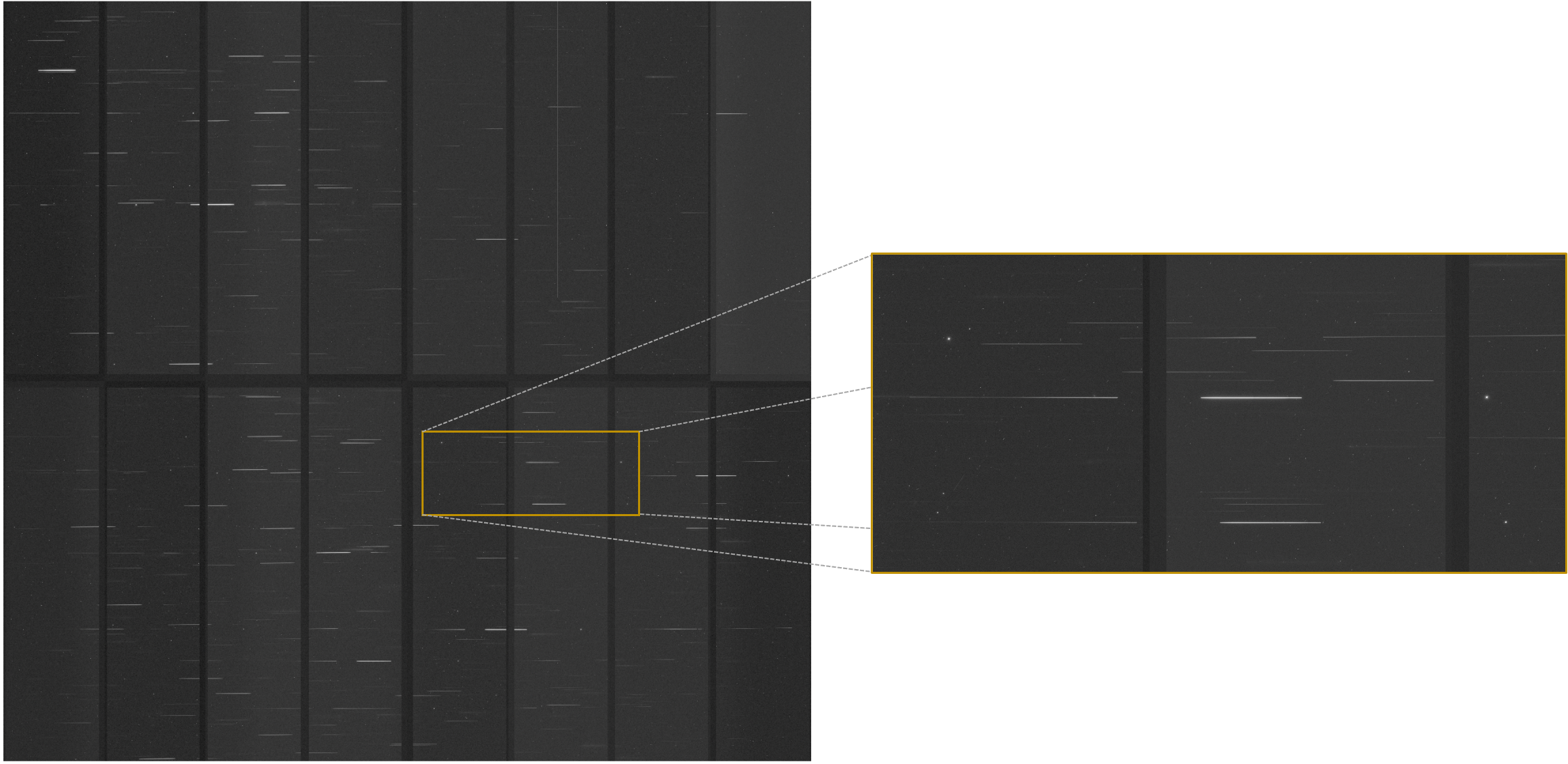}
    \caption{The log-scaled simulated image of slitless spectrum in $GV$-band, with an inset zoom-in highlighting the 0th, 1st, and 2nd order images of the slitless spectrum.}
    \label{fig:csstsim_02}
\end{figure*}

\begin{table}[h!]
    \centering
    \begin{tabular}{|l|c|c|}
        \hline
        Sky field (Ra, Dec) & \# of exposures & Volume (TB) \\ 
        \hline
        W1 (\  $30^\circ$, $-48^\circ$) & 226 & 2.3 \\
        W2 ($170^\circ$, $-23^\circ$) & 228 & 2.3 \\
        W3 ($180^\circ$, \ \ $60^\circ$) & 261 & 2.7 \\
        W4 ($240^\circ$, \ \ $30^\circ$) & 266 & 2.7 \\
        W5 ($300^\circ$, $-60^\circ$)& 240 & 2.5 \\
        \hline
        Total (25 $\deg^2$) & 1,221 & 12.5 \\
        \hline
    \end{tabular}
    \caption{Summary of recent data products simulated in C9 runs.} 
    \label{tab:C9data}
\end{table}

In practice, the CSST scientific data processing and analysis system adopts a semi-annual simulation cycle. The latest released dataset, corresponding to the Cycle-9 (C9) phase, includes simulated observations of 5 sky regions at different Galactic latitudes ( Fig.~\ref{fig:skysimulated}), specifically designed to evaluate the impact of stray light on various scientific analyses. Each region spans 5 $\deg^2$, yielding a total sky coverage of 25 $\deg^2$. To mitigate the impact of intrinsic galaxy ellipticity in the development of the weak lensing shear measurement pipeline, a complementary dataset was produced by rotating galaxy shapes by $90^\circ$, effectively doubling the coverage to 50 $\deg^2$. These simulations were completed at the National Supercomputing Center in Wuzhen, consuming a total of $\sim 300,000$ CPU-hours. The number of exposures and total data volume for each field are summarized in Tab.\ref{tab:C9data}. 

These datasets consist of raw imaging data from 7 photometric filters ($NUV$, $u$, $g$, $r$, $i$, $z$, $y$) and 3 spectroscopic gratings ($GU$, $GV$, $GI$), totaling 73,260 detector files, each with a resolution of $\rm 9K \times 9K$. As examples,  Fig.~\ref{fig:csstsim_24} and  \ref{fig:csstsim_02} show the log-scaled simulated image in $i$-band and slitless spectrum in $GV$-band, respectively. All the data have been undergone processing to produce calibrated images and catalogs, facilitated by the pipeline specifically developed for the CSST survey. 

\begin{figure*}
    \centering
    \includegraphics[width=0.95\linewidth]{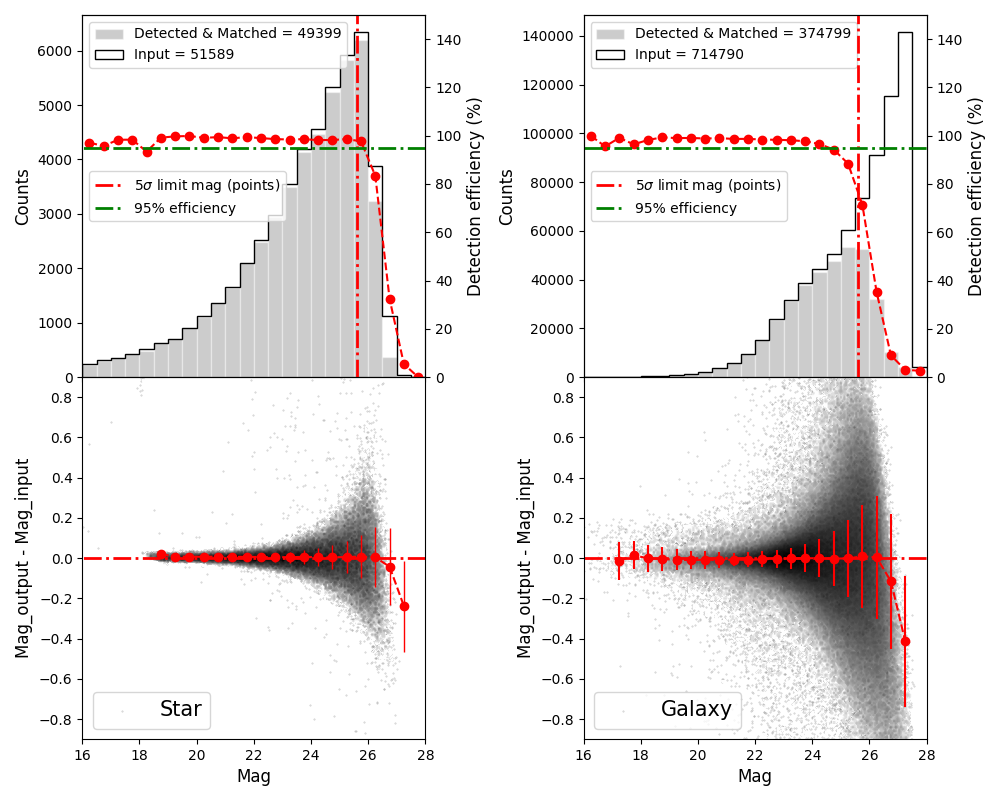}
    \caption{The comparisons between input and output catalogs for the magnitudes and brightness distributions.}
    \label{fig:magnitude_comparisions}
\end{figure*}

To ascertain the realism of the simulated noise, we conducted a comparison between the output results and the input catalogs. Fig.~\ref{fig:magnitude_comparisions} illustrates these comparisons for the magnitudes and brightness distributions, as derived from 10 randomly selected images/catalogs in the $r$ band. The input catalog was categorized into 51589 stars and 714790 galaxies, while the output magnitudes were determined using PSF photometry for stars and Kron photometry \citep{Kron1980}  for galaxies. The output magnitudes closely matched the input values, with deviations attributable to instrumental noise showing no significant systematic bias. As depicted in Fig.~\ref{fig:magnitude_comparisions}, over 95\% of the stars/point sources with a signal-to-noise ratio greater than 5 \citep{Zhan2011} were successfully detected. Additionally, the galaxies demonstrated a consistent detection level with approximately 1 magnitude brighter due to their extended profiles.

\begin{figure}
    \centering
    \includegraphics[width=0.75\textwidth]{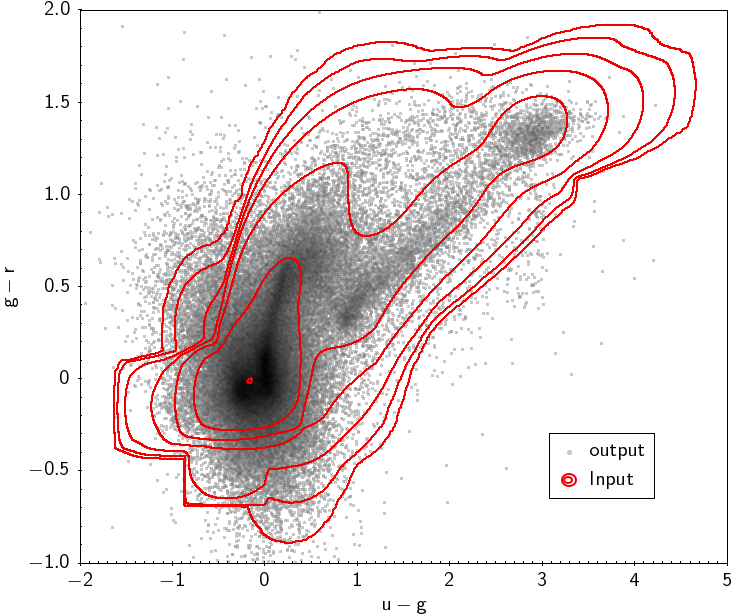}
    \caption{The u-g versus g-r diagram between the input (red contour lines) and output (black points) data.}
    \label{fig:color_color_diagram}
\end{figure}

We further examined the color-color diagrams to ensure consistency between the input and output data. The simulation dataset was partitioned into 366 HEALPix$^{[8]}$ \footnotetext{[8] \url{healpix.jpl.nasa.gov} } bricks, with a resolution of NSIDE = 256. Inside each brick, individual exposure catalogs were amalgamated into a comprehensive merging catalog that included 7-band data for every celestial source. As depicted in Fig.~\ref{fig:color_color_diagram}, we present a comparison of the u-g versus g-r diagram between the input data and a selection of 10 random merging catalogs. It is evident that the input color-color diagram closely aligns with the input catalog, and the outcomes are in agreement with corresponding plots derived from SDSS and DECALS datasets.

\section{Summary and Outlook}
\label{sect:summary}
In this work, we introduce the simulation pipeline for the Chinese Space Station Telescope (CSST), which enables the generation of high-fidelity synthetic observations for both imaging and slitless spectroscopic surveys. Our simulation framework, built on the GalSim package and incorporating detailed modeling of astronomical sources, instrumental effects, and observational conditions, is capable reproducing key observational characteristics of CSST  SCam. 

We also provide an overview of the synthetic observations produced by our simulation pipeline in the Cycle-9 data release. This dataset contains multiband imaging and spectrum observations for 2,442 exposures that covers 50 square degrees of the sky ( Fig.~\ref{fig:skysimulated}). The Cycle-9 simulation products support comprehensive testing and validation of the current CSST data processing pipelines. By comparing synthetic data with input catalogs, we achieve over $95\%$ detection efficiency for sources with a signal-to-noise ratio (SNR) $>$ 5, confirming the pipeline's capability to accurately reproduce expected magnitude and color distributions.

The CSST simulation pipeline is a powerful tool for assessing instrument performance, refining data processing algorithms, and supporting science case development for the mission. The datasets generated in this work will serve as critical testbeds for upcoming CSST data processing pipelines and scientific analyses. Looking ahead, future developments will focus on refining the simulation suite by incorporating updated CSST instrumental specifications and additional calibration data from laboratory tests. Additionally, the forthcoming Cycle-10 simulation release will expand the dataset to larger sky areas and incorporate new calibration strategies. Furthermore, the public release of this simulation framework will allow the broader astronomical community to contribute to its refinement and leverage its capabilities for a wide range of scientific investigations.

\normalem
\begin{acknowledgements}
This work is supported by the project of the CSST scientific data processing and analysis system of the China Manned Space Project. We thank the working group of CSST-`JiuTian' to make the cosmological simulations available. This work is supported by the China Manned Space Program with grant No. CMS-CSST-2025-A20, CMS-CSST-2025-A05, CMSCSST-2021-A03 and the National Natural Science Foundation of China (NSFC) No. 11903082, 12003001.

\end{acknowledgements}
  
\bibliographystyle{raa}
\bibliography{bibtex}

\begin{thebibliography}{54}
\providecommand\natexlab[1]{#1}
\providecommand\JournalTitle[1]{#1}

\bibitem[{Allard} {et~al.}(2011)]{Allard2011ASPC..448...91A}
{Allard}, F., {Homeier}, D., \& {Freytag}, B. 2011, in Astronomical Society of the Pacific Conference Series, Vol. 448, 16th Cambridge Workshop on Cool Stars, Stellar Systems, and the Sun, ed. C.~{Johns-Krull}, M.~K. {Browning}, \& A.~A. {West}, 91

\bibitem[Anderson \& Bedin(2010)]{Anderson_2010}
Anderson, J., \& Bedin, L.~R. 2010, Publications of the Astronomical Society of the Pacific, 122, 1035

\bibitem[{Antilogus} {et~al.}(2014)]{2014JInst...9C3048A}
{Antilogus}, P., {Astier}, P., {Doherty}, P., {Guyonnet}, A., \& {Regnault}, N. 2014, Journal of Instrumentation, 9, C03048

\bibitem[{Astier} \& {Regnault}(2023)]{2023A&A...670A.118A}
{Astier}, P., \& {Regnault}, N. 2023, \aap, 670, A118

\bibitem[{Broughton} {et~al.}(2024)]{2024PASP..136d5003B}
{Broughton}, A., {Utsumi}, Y., {Plazas Malag{\'o}n}, A.~A., {et~al.} 2024, \pasp, 136, 045003

\bibitem[{Calabretta} \& {Greisen}(2002)]{2002A&A...395.1077C}
{Calabretta}, M.~R., \& {Greisen}, E.~W. 2002, \aap, 395, 1077

\bibitem[{Camps} \& {Baes}(2015)]{Camps+2015A&C_SKIRT}
{Camps}, P., \& {Baes}, M. 2015, Astronomy and Computing, 9, 20

\bibitem[{Chen} {et~al.}(2023)]{Chen2023}
{Chen}, Y., {Fu}, X., {Liu}, C., {et~al.} 2023, Science China Physics, Mechanics, and Astronomy, 66, 119511

\bibitem[{Euclid Collaboration} {et~al.}(2022)]{Euclid_prepare}
{Euclid Collaboration}, {Scaramella}, R., {Amiaux}, J., {et~al.} 2022, \aap, 662, A112

\bibitem[{Fu} {et~al.}(2013)]{Fu+2013MNRAS}
{Fu}, J., {Kauffmann}, G., {Huang}, M.-l., {et~al.} 2013, \mnras, 434, 1531

\bibitem[{Gentile} {et~al.}(2013)]{2013A&A...549A...1G}
{Gentile}, M., {Courbin}, F., \& {Meylan}, G. 2013, \aap, 549, A1

\bibitem[{Gong} {et~al.}(2019)]{Gong2019}
{Gong}, Y., {Liu}, X., {Cao}, Y., {et~al.} 2019, \apj, 883, 203

\bibitem[{Gong} {et~al.}(2025)]{Gong_2025}
{Gong}, Y., {Miao}, H., {Zhou}, X., {et~al.} 2025, arXiv e-prints, arXiv:2501.15023

\bibitem[{Greisen} \& {Calabretta}(2002)]{2002A&A...395.1061G}
{Greisen}, E.~W., \& {Calabretta}, M.~R. 2002, \aap, 395, 1061

\bibitem[{Guo} {et~al.}(2013)]{Guo+2013MNRAS}
{Guo}, Q., {White}, S., {Angulo}, R.~E., {et~al.} 2013, \mnras, 428, 1351

\bibitem[Hall(1952)]{PhysRev.87.387}
Hall, R.~N. 1952, Phys. Rev., 87, 387

\bibitem[{Han} {et~al.}(2025)]{Jiutian_2025arXiv250321368H}
{Han}, J., {Li}, M., {Jiang}, W., {et~al.} 2025, arXiv e-prints, arXiv:2503.21368

\bibitem[{Hauschildt} {et~al.}(1997)]{Hauschildt1997ApJ...483..390H}
{Hauschildt}, P.~H., {Baron}, E., \& {Allard}, F. 1997, \apj, 483, 390

\bibitem[{Henriques} {et~al.}(2015)]{Henriques+2015MNRAS}
{Henriques}, B. M.~B., {White}, S. D.~M., {Thomas}, P.~A., {et~al.} 2015, \mnras, 451, 2663

\bibitem[Israel {et~al.}(2015)]{8200689}
Israel, H., Massey, R., Prod'homme, T., {et~al.} 2015, Monthly Notices of the Royal Astronomical Society, 453, 561

\bibitem[{Klioner}(2003)]{Klioner2003AJ}
{Klioner}, S.~A. 2003, The Astronomical Journal, 125, 1580

\bibitem[{Kron}(1980)]{Kron1980}
{Kron}, R.~G. 1980, \apjs, 43, 305

\bibitem[{K{\"u}mmel} {et~al.}(2009)]{2009PASP..121...59K}
{K{\"u}mmel}, M., {Walsh}, J.~R., {Pirzkal}, N., {Kuntschner}, H., \& {Pasquali}, A. 2009, \pasp, 121, 59

\bibitem[{Lage} {et~al.}(2021)]{2021JAP...130p4502L}
{Lage}, C., {Bradshaw}, A., {Anthony Tyson}, J., \& {LSST Dark Energy Science Collaboration}. 2021, Journal of Applied Physics, 130, 164502

\bibitem[{Lage} {et~al.}(2017)]{2017arXiv170305823L}
{Lage}, C., {Bradshaw}, A., \& {Tyson}, J.~A. 2017, arXiv e-prints, arXiv:1703.05823

\bibitem[{Laureijs} {et~al.}(2011)]{Euclid_Definition}
{Laureijs}, R., {Amiaux}, J., {Arduini}, S., {et~al.} 2011, arXiv e-prints, arXiv:1110.3193

\bibitem[Leinert {et~al.}(1998)]{Leinert1998}
Leinert, C., Bowyer, S., Haikala, L.~K., {et~al.} 1998, Astron. Astrophys. Suppl. Ser., 127, 1

\bibitem[{Luo} {et~al.}(2016)]{Luo+2016MNRAS}
{Luo}, Y., {Kang}, X., {Kauffmann}, G., \& {Fu}, J. 2016, \mnras, 458, 366

\bibitem[Massey {et~al.}(2010)]{8223496}
Massey, R., Stoughton, C., Leauthaud, A., {et~al.} 2010, Monthly Notices of the Royal Astronomical Society, 401, 371

\bibitem[{Massey} {et~al.}(2013)]{2013MNRAS.429..661M}
{Massey}, R., {Hoekstra}, H., {Kitching}, T., {et~al.} 2013, \mnras, 429, 661

\bibitem[Massey {et~al.}(2014)]{8184054}
Massey, R., Schrabback, T., Cordes, O., {et~al.} 2014, Monthly Notices of the Royal Astronomical Society, 439, 887

\bibitem[{McGreer} {et~al.}(2021)]{McGreer+2021ascl.soft06008M}
{McGreer}, I., {Moustakas}, J., \& {Schindler}, J. 2021, {simqso: Simulated quasar spectra generator}, Astrophysics Source Code Library, record ascl:2106.008

\bibitem[{O'Donnell}(1994)]{O'Donnell1994}
{O'Donnell}, J.~E. 1994, \apj, 422, 158

\bibitem[{Planck Collaboration} {et~al.}(2014)]{Planck2014}
{Planck Collaboration}, {Abergel, A.}, {Ade, P. A. R.}, {et~al.} 2014, \aap, 571, A11

\bibitem[{Planck Collaboration} {et~al.}(2020{\natexlab{a}})]{Planck_2018}
{Planck Collaboration}, {Aghanim}, N., {Akrami}, Y., {et~al.} 2020{\natexlab{a}}, \aap, 641, A6

\bibitem[{Planck Collaboration} {et~al.}(2020{\natexlab{b}})]{Aghanim+2020A&A_Planck}
{Planck Collaboration}, {Aghanim}, N., {Akrami}, Y., {et~al.} 2020{\natexlab{b}}, \aap, 641, A6

\bibitem[{Qiu} \& {Kang}(2022)]{Qiu+2022ApJ}
{Qiu}, Y., \& {Kang}, X. 2022, \apj, 930, 66

\bibitem[Riess(2002)]{Riess2002}
Riess, A. 2002, Instrument Science Report ACS 2002-07, 13 pages

\bibitem[{Riess} {et~al.}(1998)]{Riess_1998}
{Riess}, A.~G., {Filippenko}, A.~V., {Challis}, P., {et~al.} 1998, \aj, 116, 1009

\bibitem[{Robin} {et~al.}(2003)]{Robin2003A&A...409..523R}
{Robin}, A.~C., {Reyl{\'e}}, C., {Derri{\`e}re}, S., \& {Picaud}, S. 2003, \aap, 409, 523

\bibitem[{Ross} {et~al.}(2013)]{Ross+2013ApJ}
{Ross}, N.~P., {McGreer}, I.~D., {White}, M., {et~al.} 2013, \apj, 773, 14

\bibitem[{Rowe} {et~al.}(2015)]{Rowe2015}
{Rowe}, B.~T.~P., {Jarvis}, M., {Mandelbaum}, R., {et~al.} 2015, Astronomy and Computing, 10, 121

\bibitem[{Sharma} {et~al.}(2011)]{Galaxia2011ApJ...730....3S}
{Sharma}, S., {Bland-Hawthorn}, J., {Johnston}, K.~V., \& {Binney}, J. 2011, \apj, 730, 3

\bibitem[Shockley \& Read(1952)]{PhysRev.87.835}
Shockley, W., \& Read, W.~T. 1952, Phys. Rev., 87, 835

\bibitem[Short {et~al.}(2013)]{8185613}
Short, A., Crowley, C., de~Bruijne, J. H.~J., \& Prod'homme, T. 2013, Monthly Notices of the Royal Astronomical Society, 430, 3078

\bibitem[{Spergel} {et~al.}(2015)]{Roman_report}
{Spergel}, D., {Gehrels}, N., {Baltay}, C., {et~al.} 2015, arXiv e-prints, arXiv:1503.03757

\bibitem[{Springel} {et~al.}(2001)]{Springel+2001MNRAS_subfind}
{Springel}, V., {White}, S. D.~M., {Tormen}, G., \& {Kauffmann}, G. 2001, \mnras, 328, 726

\bibitem[{Wei} {et~al.}(2018{\natexlab{a}})]{wei+2018MNRAS}
{Wei}, C., {Li}, G., {Kang}, X., {et~al.} 2018{\natexlab{a}}, \mnras, 478, 2987

\bibitem[{Wei} {et~al.}(2018{\natexlab{b}})]{Wei+2018ApJ}
{Wei}, C., {Li}, G., {Kang}, X., {et~al.} 2018{\natexlab{b}}, \apj, 853, 25

\bibitem[{Zhan}(2011)]{Zhan2011}
{Zhan}, H. 2011, Scientia Sinica Physica, Mechanica \& Astronomica, 41, 1441

\bibitem[Zhan(2021)]{Zhan2021}
Zhan, H. 2021, Chinese Science Bulletin, 66, 1290

\bibitem[{Zhang} {et~al.}(2019)]{2019ApJ...875...48Z}
{Zhang}, J., {Dong}, F., {Li}, H., {et~al.} 2019, \apj, 875, 48

\bibitem[{Zhang} {et~al.}(2025)]{zhangxin2025}
{Zhang}, X., {Fang}, Y.-D., {Wei}, C.-L., {et~al.} 2025, RAA, in press

\bibitem[{Zhang} \& {Yang}(2019)]{Zhang+2019RAA}
{Zhang}, Y.-C., \& {Yang}, X.-H. 2019, Research in Astronomy and Astrophysics, 19, 006

\end{thebibliography}

\appendix

\section{Customizable Interfaces for Input Truth Catalogs}\label{sect:customizable_catalog}
The simulation suite accepts lists of \texttt{MockObject} instances as its primary form of input. As a result, any object from the "True Universe" must first be parsed and instantiated into corresponding \texttt{MockObject} instances. This step ensures compatibility between external data and the simulation's internal structure.

Furthermore, the simulation suite has been specifically designed to be open, flexible, and highly adaptable, making it suitable for a wide range of potential use cases across various fields and applications. To enable this level of flexibility, the suite includes a meta-class interface called \texttt{CatalogBase}. This interface serves as a foundational framework that allows users to easily extend or customize the process of parsing and instantiating objects from their own “True Universe”. In particular, users are required to:
\begin{itemize}
    \item Implement the constructor \texttt{\_\_init\_\_()}: This is typically used to load relevant data into memory for further processing.
    \item Implement the \texttt{\_load()} method: This method parses objects from the user’s “True Universe” and converts them into \texttt{MockObject} instances, storing them in a list.
    \item Implement the \texttt{load\_sed()} method: This method is responsible for loading and parsing the SED for each object into an \texttt{Astropy.Table}.
    \item Implement the \texttt{load\_norm\_filt()} method (if applicable): If an object’s magnitudes in the user’s “True Universe” are not already in the CSST magnitude system, this method must be implemented to load the corresponding filter responses. This enables the simulation suite to perform the necessary magnitude conversions.
\end{itemize}
By implementing these methods, users can seamlessly integrate their own data into the simulation framework while maintaining full compatibility with its internal processes.

Alongside \texttt{CatalogBase}, the suite includes a range of tools for tasks such as synthetic magnitudes, astrometric and distance calculations, and Milky Way extinction estimations, among others. These tools enable users to customize and optimize the parsing process according to their specific needs, making the simulation adaptable to diverse datasets and operational requirements.

\section{Simulation Intermediate Layers} \label{sect:simulation_layers}
As of the writing of this paper, 16 intermediate layers have been implemented. They are summarized as follows (The bold text in parentheses represents the corresponding layer keywords and relevant parameters):
\begin{enumerate}
    \setlength{\itemsep}{2pt}
    \item \textbf{Scientific objects exposure (\texttt{scie\_obs})}: Parameterized by the exposure time (\textbf{\texttt{exptime}}), the accumulated photon distributions from objects in the truth catalogs are modeled and then rendered onto a CSST chip.  On/off toggles are provided for modeling the reversal of flat fielding correction (\textbf{\texttt{flat\_fielding}}), vignetting due to shutter movements (\textbf{\texttt{shutter\_effect}}), as well as the positional displacements and shape changes due to field distortion effect (\textbf{\texttt{field\_dist}}). Further details on modeling  and field distortion effect for objects  are provided in section~\ref{sect:object_modeling} and section~\ref{sect:field_distortion}.
    
    \item \textbf{Sky background and stray light (\texttt{sky\_background})}: Parameterized by the exposure time (\textbf{\texttt{exptime}}), and based on the corresponding band-pass, accumulated photons from the background sky are calculated and added to the chip image. There is an option (\textbf{\texttt{enable\_straylight\_model}}) to enable the modeling of stray lights, which currently includes estimations of earth shine, zodiacal light, and starlight outside the FOV. Similar to the scientific objects exposure layer, \textbf{\texttt{flat\_fielding}} and \textbf{\texttt{shutter\_effect}} switches are also provided to account for the photon loss due to vignetting and shutter movements. 
    
    \item \textbf{Pixel Response Non-uniformity (\texttt{PRNU\_effect})}: A weight map of pixel response non-uniformity is applied to the corresponding chip image. Currently, simple Gaussian models are used to generate the weight map for each chip, but these are expected to be replaced by lab-measured values.
    \item \textbf{Cosmic ray accumulation (\texttt{cosmic\_rays})}: Parameterized by the exposure time (\textbf{\texttt{exptime}}), this layer creates a mock cosmic ray map, which is then added pixel-wisely to the chip image. The charge diffusion effect has been simulated in this layer. An option (\textbf{\texttt{save\_cosmic\_image}}) is available to save this intermediate map.
    \item \textbf{Poisson noise and dark current (\texttt{poisson\_and\_dark})}: Add Poisson noise to each pixel based on its value. This step also add dark current according to the exposure time (\textbf{\texttt{exptime}}). This step is usually executed right after all photoelectrons have been accumulated.
    \item \textbf{Brighter fatter effect (\texttt{brighter\_fatter})}: Apply the phenomenological model described in section~\ref{sect:bright_fatter} to simulate brighter-fatter effect on the charge distribution. 
    \item \textbf{Hot pixels, dead pixels, and invalid columns (\texttt{detector\_defects})}: Add any combination of three types of detector defects (\textbf{\texttt{hot\_pixels}}, \textbf{\texttt{dead\_pixels}}, and \textbf{\texttt{bad\_columns}}) at random locations on the image.
    \item \textbf{Charge-to-voltage conversion non-linearity (\texttt{nonlinearity})}: This layer simulates the non-linear behavior of the charge-to-voltage conversion process. It applies a third-order polynomial transformation to the image, capturing the inherent non-linear relationship between the accumulated charge and the resulting voltage signal.
    \item \textbf{Pixel saturation and bleeding (\texttt{blooming})}: This layer mimics the overflow of charges from highly saturated pixels into adjacent regions. It models the blooming effect observed in saturated pixels by simulating the movement of excess charges along both the x and y axes.
    \item \textbf{Charge transfer inefficiency (\texttt{CTE\_effect})}: As described in section~\ref{sect:CTI}, this layer simulates the presence of residual charges that result from the incomplete or inefficient transfer of charges during the readout process.
    \item \textbf{Simulating pre-scan and over-scan sections (\texttt{prescan\_overscan})}: Simulates and adds the pre-scan and over-scan areas of each detector to the image.
    \item \textbf{Adding bias (\texttt{bias})}: Adds bias to the whole image or each of the 16 readout channels (\textbf{\texttt{bias\_16channel}}).
    \item \textbf{Simulating readout noise (\texttt{readout\_noise})}: Adds a Gaussian noise due to readout process.
    \item \textbf{Applying gain (\texttt{gain})}: Apply the amplification gain to the whole image or each of the 16 readout channels (\textbf{\texttt{gain\_16channel}}).
    \item \textbf{Outputting the digitized data product (\texttt{quantization\_and\_output})}: This layer digitizes the image to 16 bits, appends the corresponding FITS headers and checksum, and outputs the data product in the format specified by the CSST level-0 data definition.
    \item \textbf{LED illuminations (\texttt{led\_calib\_model})}: Simulates LED-based flat-field calibration exposures. A list of strings (\textbf{\texttt{LED\_TYPE}}) is required to define the types and sequence of switching on the LEDs. Additionally, a list of floats (\textbf{\texttt{LED\_TIME}}) specifies the corresponding illumination times.
\end{enumerate}

\section{the Extinction of the Milky Way } \label{sect:milky_way_extinction}

Interstellar extinction caused by dust and gas within the Milky Way affects the observed fluxes and SEDs of astronomical objects by absorbing and scattering light. This extinction is wavelength-dependent, resulting in both a reduction in overall brightness and a reddening of the source’s spectrum. Accurately accounting for this effect is essential for producing realistic synthetic observations. 

In our simulator, the extinction of the Milky Way  is applied to galaxies by scaling the redshifted and reddened SED fluxes:
\[f_{\rm obs}(\lambda) = f(\lambda) \times 10^{-0.4A(\lambda)}\]
Where
\(A(\lambda)/A(V)\) follows \citealt{O'Donnell1994} extinction law. Here \(A(V)=E(B-V)\times R_V\) and \(R_V=3.1\). The colour excess \(E(B-V)\) is obtained from the Planck dust maps (\citealt{Planck2014}).

For stars, we followed the method described in Section 3.2 of  \cite{Chen2023}. The \(E(B-V)\) values are retrieved from a combination of Planck dust maps:
\begin{multline*}
E(B-V)_{\rm distant}= 
\begin{cases}
E(B-V)_{\chi\rm{gal}} & E(B-V)_{\chi\rm{gal}} < 0.3\ \rm{mag}\\
1.49 \times 10^4 \times \tau_{353} & \rm{otherwise}
\end{cases}
\end{multline*}
Assuming an exponential distribution of dust along the LOS, \(E(B-V)_{distant}\) is then scaled based on the LOS distance of each star to get the corresponding colour excess \(E(B-V)\). Then we apply the same extinction law and scaling on the SED fluxes as for galaxies.

This extinction correction is applied directly to the observed-frame SEDs prior to flux integration and photon rendering.

\end{document}